\documentclass{aa}  
\usepackage{natbib}
\usepackage{xcolor}
\usepackage{graphicx}
\usepackage{natbib}
\bibpunct{(}{)}{;}{a}{}{,}
\usepackage{diagbox}
\usepackage{txfonts}
\usepackage{xcolor}
\usepackage{makecell}
\newcommand{\blos}{$B_{\parallel}$}

\newcommand{\Av}{$A_{{V}}$}
\newcommand{\bperp}{$B_{\perp}$}
\usepackage{scalerel}
\newcommand{\HII}{H\protect\scaleto{$II$}{1.2ex}}
\newcommand{\HI}{H\protect\scaleto{$I$}{1.2ex}}
\newcommand{\vcoHi}{$\varv_{\text{CO - H\protect\scaleto{$I$}{1.2ex}}}$}
\newcommand{\vco}{$\varv_{\text{CO, LSR}}$}
\newcommand{\vhi}{$\varv_{\text{H\protect\scaleto{$I$}{1.2ex}, LSR}}$}

\usepackage{subcaption}

\begin{document} 
  \title{3D magnetic field morphology of the Perseus molecular cloud}
  \author{ M. Tahani \inst{1}
          \and W. Lupypciw\inst{1,2}
          \and J. Glover\inst{1, 3}
          \and R. Plume\inst{2}
          \and J.L. West\inst{4}
          \and R. Kothes\inst{1}
          \and S. Inutsuka\inst{5}
          \and M-Y. Lee\inst{6}
          \and T. Robishaw\inst{1}
          \and L.B.G. Knee\inst{7}
          \and J.C. Brown\inst{2}
          \and Y. Doi\inst{8}
          \and I.A. Grenier\inst{9}
          \and M. Haverkorn\inst{10}
         }

   \institute{Dominion Radio Astrophysical Observatory, Herzberg Astronomy and Astrophysics Research Centre, National Research Council Canada, P. O. Box 248, Penticton, BC V2A 6J9 Canada\\
   \email{mehrnoosh.tahani@nrc.ca}
   \and Department of Physics \& Astronomy, University of Calgary, Calgary, Alberta T2N 1N4, Canada
  \and Department of Physics \& Astronomy, University of Victoria, Victoria, British Columbia V8P 5C2, Canada
    \and
    Dunlap Institute for Astronomy and Astrophysics University of Toronto, Toronto, ON M5S 3H4, Canada
    \and 
    Department of Physics, Graduate School of Science, Nagoya University, Furo-cho, Chikusa-ku, Nagoya 464-8602, Japan
    \and
    Korea Astronomy and Space Science Institute, 776 Daedeok-daero, 34055 Daejeon, Republic of Korea 
    \and 
    Herzberg Astronomy and Astrophysics Research Centre, National Research Council Canada, 5071 West Saanich Road, Victoria BC V9E 2E7, Canada
    \and 
    Department of Earth Science and Astronomy, Graduate School of Arts and Sciences, The University of Tokyo, 3-8-1 Komaba, Meguro, Tokyo 153-8902, Japan
    \and
    Universit\'{e} de Paris and Universit\'{e} Paris Saclay, CEA, CNRS, AIM, F-91190 Gif-sur-Yvette, France
    \and 
    Department of Astrophysics/IMAPP, Radboud University, P.O. Box 9010,
6500 GL Nijmegen, The Netherlands
             }

   \date{Received Date; accepted Date}

\titlerunning{The Perseus cloud's 3D magnetic field}
\authorrunning{M. Tahani et al.}
 
\abstract
{Despite recent observational and theoretical advances in mapping the magnetic fields associated with molecular clouds,  their three-dimensional (3D) morphology remains unresolved.  Multi-wavelength and multi-scale observations will allow us to paint a comprehensive picture of the magnetic fields of these star-forming regions.}
{We reconstruct the 3D magnetic field morphology associated with the Perseus molecular cloud and compare it with predictions of cloud-formation models. These cloud-formation models predict a bending of magnetic fields associated with filamentary molecular clouds. We compare the orientation and direction of this field bending with our 3D magnetic field view of the Perseus cloud. }
{We use previous line-of-sight and plane-of-sky magnetic field observations, as well as Galactic magnetic field models, to reconstruct the complete 3D magnetic field vectors and morphology associated with the Perseus cloud. }
{We approximate the 3D magnetic field morphology of the  cloud as a concave arc that points in the decreasing longitude direction in the plane of the sky (from our point of view). This field morphology preserves a memory of the Galactic magnetic field. In order to compare this morphology to cloud-formation model predictions, we assume that the cloud retains a memory of its most recent interaction. Incorporating velocity observations, we find that the line-of-sight magnetic field observations are consistent with predictions of shock-cloud-interaction models.}
{To our knowledge, this is the first time that the 3D magnetic fields of a molecular cloud have been reconstructed. We find the 3D magnetic field morphology of the Perseus cloud to be consistent with the predictions of the shock-cloud-interaction model, which describes the formation mechanism of filamentary molecular clouds. 
}

   
\maketitle
%
\section{Introduction}
\label{intro}

Molecular clouds, where stars are formed, are often shaped as elongated filamentary structures or filaments~\cite[e.g.,][]{Molinari2010, Andre2010, Andreetal2014, Arzoumanian2011}. Within these structures, magnetic fields play an important role in the star-formation process~\citep[e.g.,][]{MckeeOstriker2007, HennebelleFalgarone2012, SeifriedWalch2015, HennebelleInutsuka2019, PudritzRay2019, PattleFissel2019}. Determining the three-dimensional (3D) magnetic field morphology~\citep{Houdeetal2004, LiHoude2008} of these star-forming regions will enable us to characterize their role in the process. Recent studies~\citep[e.g.,][]{Chenetal2018, Tahanietal2019, YueLazarian2020} have investigated these 3D fields using different techniques. 

To fully probe 3D magnetic fields, observations of both the plane-of-sky  and the line-of-sight components of magnetic fields (\bperp\ and \blos , respectively) are necessary. Recent observations have made significant progress in mapping the \bperp\ of molecular clouds~\citep[e.g.,][]{PlanckXXXV, Fisseletal2016, Pattleetal2019, Doietal2020}. \citet{PlanckXXXII,  PlanckXXXV} showed that \bperp\ lines tend to be perpendicular to high column density ($> 10^{21.7}$\,cm$^{-2}$) filamentary structures and parallel to lower column density ones. Simulations show that magnetic fields perpendicular to the filaments allow for greater mass accumulations and result in denser filaments \cite[e.g.,][]{Inoueetal2018, HennebelleInutsuka2019}.

\begin{figure}[htbp]
\centering
\includegraphics[scale=0.3, trim={0cm 0cm 0cm 0cm},clip]{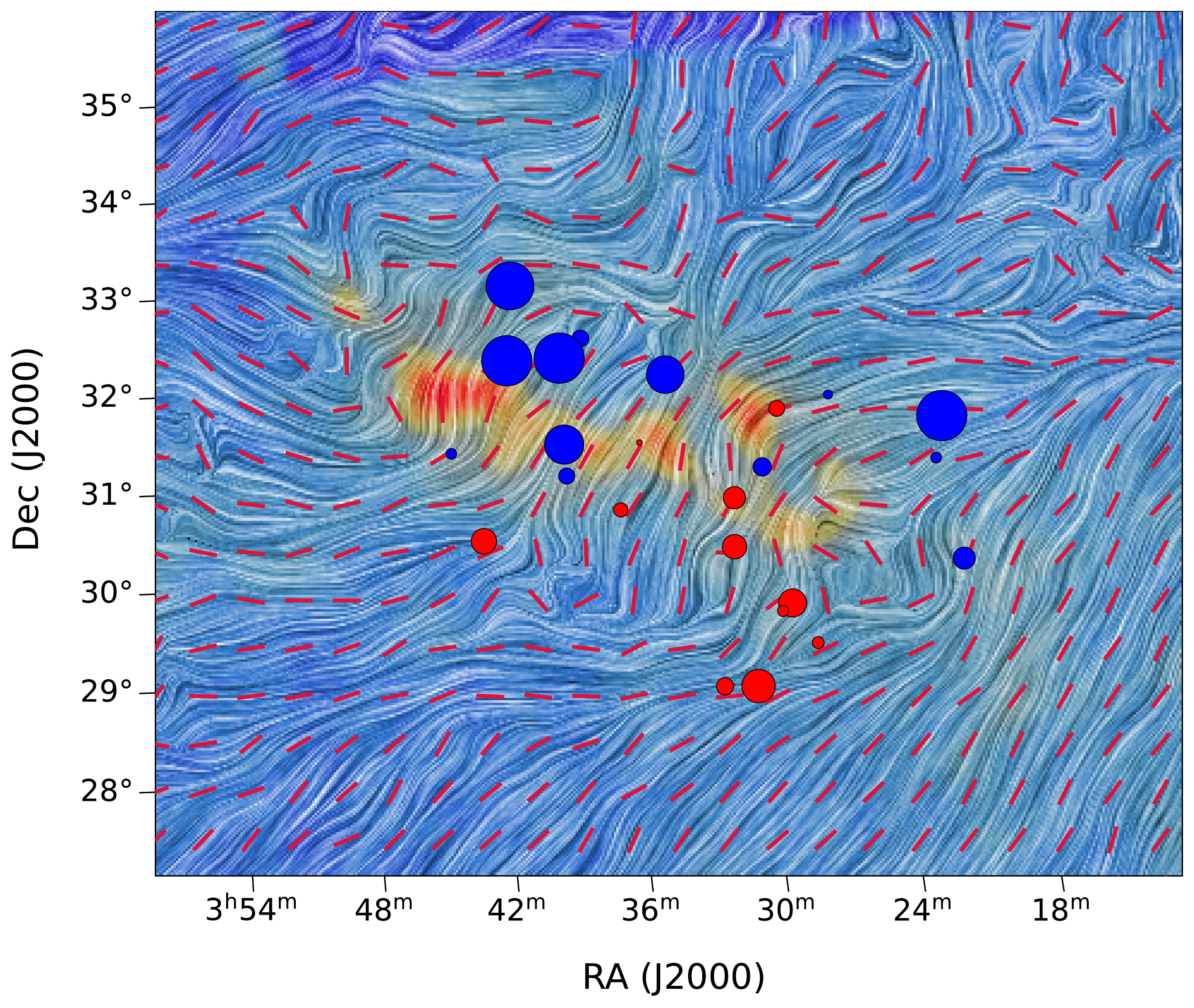}
\caption{The observed magnetic fields in the vicinity of the Perseus molecular cloud. The \bperp\ lines were observed by the Planck Space Observatory and the \blos\ were obtained by~\citet{Tahanietal2018}. The red and drapery lines show the \bperp , and the blue [red] circles show magnetic fields toward [away from] us. The size of the circles represents the strength of the magnetic fields. The background color image is the visual extinction map obtained by \citet{Kainulainenetal2009}.}
\label{fig:PerseusBlosBpos}
\end{figure}

A recent study by \citet{Tahanietal2018} developed a new technique based on Faraday rotation measurements to map the \blos\ associated with molecular clouds\footnote{Recently updated code for determining \blos\ is available at \url{https://github.com/MehrnooshTahani/MappingBLOS_MolecularClouds}}. They found that in some molecular clouds, including the Perseus cloud, the \blos\ direction reverses from one side of these filamentary clouds to the other side (perpendicular to the cloud's long axis), as shown in Figure~\ref{fig:PerseusBlosBpos}. In this figure, the background color image shows the visual extinction map (in units of magnitude of visual extinction or \Av ) of the Perseus  cloud provided by \citet{Kainulainenetal2009}, where they obtained near-infrared dust extinction maps using the 2MASS data archive and the NICEST \citep{Lombardi2009Nicest} color excess mapping technique. The blue [red] circles show magnetic fields toward [away from] us. The red and drapery lines~\citep[made using the line integration convolution technique\footnote{\url{https://pypi.org/project/licpy/}},][]{Cabral1993LIC} show the \bperp\ observed by Planck. \citet{PlanckXXXV} and \citet{Soler2019} show that \bperp\ is mostly perpendicular to the Perseus cloud. 

An arc-shaped\footnote{previously called bow-shaped~\citep{Tahanietal2019}; pronounced /b\={o}/ as in rainbow or bow and arrow} magnetic field morphology has been proposed to explain this \blos\ reversal across the clouds. For example, \citet{Heiles1997} suggested that an arc-shaped magnetic field morphology should be present associated with the Orion A cloud, due to recurrent shocks by nearby supernovae in the Orion-Eridanus bubble.  \citet{Tahanietal2019} showed that an arc-shaped magnetic field morphology was the most likely candidate among the magnetic morphologies that could explain a \blos\ reversal across the Orion A filamentary cloud.

\begin{figure}[htbp]
\centering
\includegraphics[scale=0.3, trim={0cm 0.cm 0cm 0.5cm},clip]{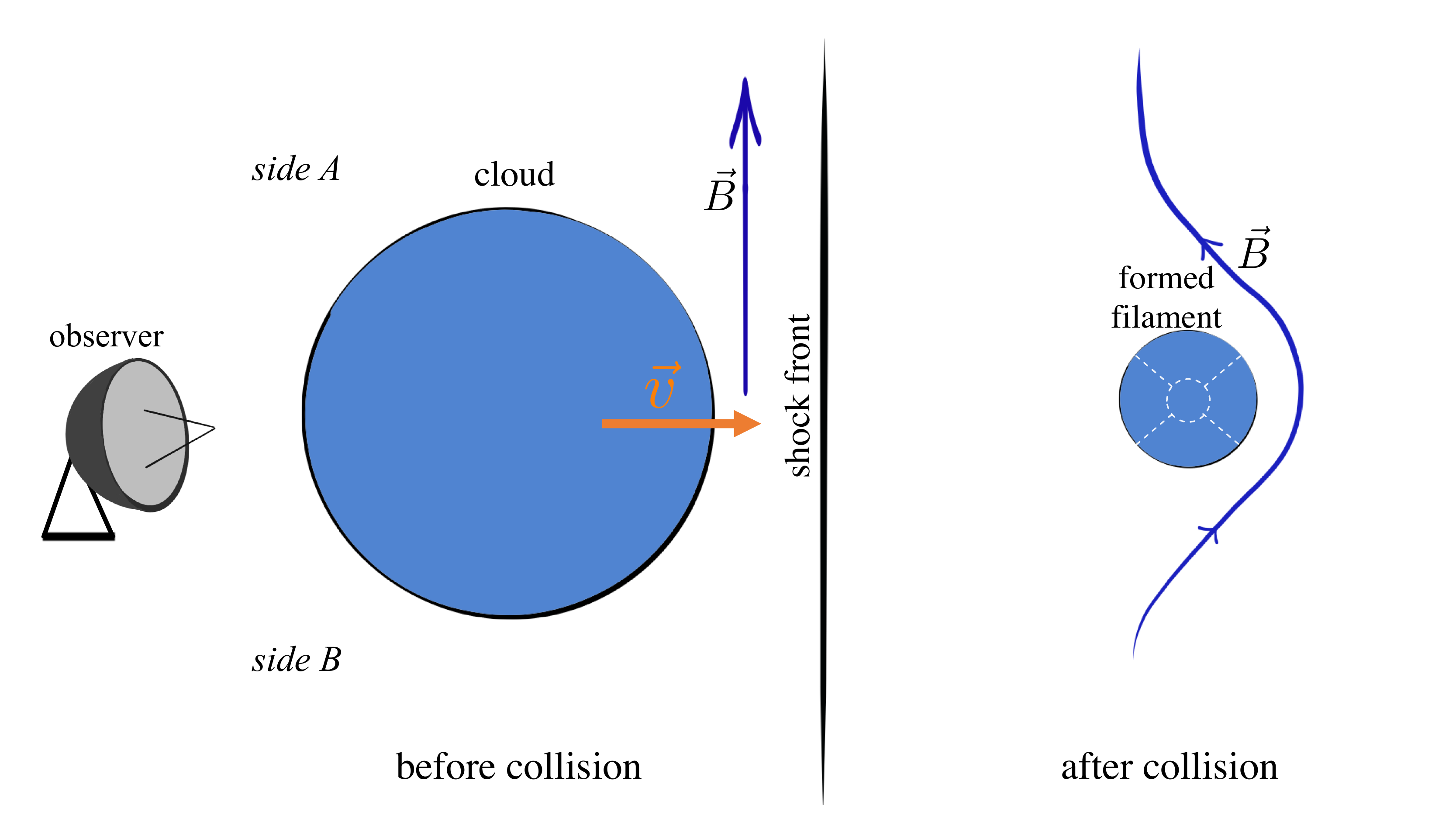}
\caption{Formation of an arc-shaped magnetic field morphology around filamentary molecular clouds as described by \citet{Inoueetal2018}. After the interaction between the cloud and shock-front, with the velocity $\vec{\varv}$ in the co-moving frame of the shock-front, a filamentary structure is formed (shown in an end-on view in the right side). $\vec{B}$ shows the direction of the initial magnetic field before the collision and the morphology of the magnetic field after the collision. The resulting field structure in this image is concave from the observer's vantage point. }
\label{fig:SCIModel}
\end{figure}

Moreover, studies of \citet{Doietal2021} and \citet{Bialyetal2021} have suggested that bubbles in the Perseus-Taurus region have formed the Perseus molecular cloud. Formation of clouds through expanding bubbles can result in an arc-shaped magnetic field morphology associated with the filamentary structures~\citep{FukuiInoue2013, Inutsukaetal2015, Inoueetal2018, Abeetal2020}. These cloud-formation studies suggest that dense filamentary molecular clouds form via interaction of a shock front and a pre-existing inhomogeneous dense cloud, 
where the magnetic fields are predicted to bend and allow for further mass accumulation, creating  even denser filaments. This results in filaments or filamentary structures with an arc-shaped magnetic field around them. 
In this scenario (shock-cloud interaction; hereafter SCI model), 
the interaction occurs between a relatively dense cloud ($\sim 10^3$\,cm$^{-3}$) and a shock wave propagating in low density gas (\HI ), as illustrated in Figure~\ref{fig:SCIModel}. 
 
In general, the shock wave propagation direction is not aligned with the mean magnetic field. Therefore, the SCI model considers a perpendicular magnetic field ($\vec{B} \perp \vec{\varv}$) as illustrated in Figure~\ref{fig:SCIModel}. 
This process is studied in detail by \citet{Abeetal2020} as a formation mechanism of filamentary structures on smaller scales (a few pc). However, the resultant geometry of the shock-cloud interaction is scale-free and should be the same on larger scales. 
Evidence of magnetic field bending due to environmental effects has been observed on large scales \cite[$\sim$ 100 pc;][]{Soleretal2018, Braccoetal2020}. 

In this study we investigate the 3D magnetic field morphology of the Perseus molecular cloud, which is actively forming a significant number of low to intermediate-mass stars
\citep{Ballyetal2008}. This cloud encompasses 10$^{4}$ solar masses \citep[M$_{\odot}$;][]{Ballyetal2008}, and has a distance of $294 \pm 10$ pc from the Sun~\citep{Zuckeretal2019}. The formation scenario proposed in the studies of \citet{Doietal2020} and \citet{Bialyetal2021} indicates a high likelihood for an arc-shaped magnetic field morphology in this region. 

We explore the direction and the shape of this arc-shaped field morphology in this study and reconstruct the most likely 3D magnetic field morphology associated with the Perseus cloud. For this purpose we incorporate the \blos\ and \bperp\ observations of \citet{Tahanietal2018} and \citet{PlanckXXXV}, as well as Galactic magnetic field models to estimate the initial magnetic field orientation and direction in this region. We then compare this 3D magnetic field morphology with cloud-formation model predictions. To this end, we use the \HI\ and CO velocity observations in this region, as well as other studies and observational data indicating interactions between the Perseus cloud and its surroundings. 

Section~\ref{sec:data} presents the observational data that are used in this study. We  discuss our approach and reconstruct the Perseus cloud's 3D magnetic field in Section~\ref{sec:3Dmorphologies}. We compare the obtained 3D magnetic field with the predictions of cloud-formation models in Section~\ref{sec:formationScenario}, using velocity observations. Finally, we provide a general discussion in Section~\ref{discussion}, followed by a summary and conclusion in Section~\ref{sec:summary}.

\section{Data used in the study}
\label{sec:data}

We incorporate estimates of the initial 3D magnetic field direction, cloud velocities, and the present-state \blos\ observations associated with the Perseus cloud. To estimate the initial magnetic field direction, we use magnetic field models of the Galaxy. To determine the cloud velocities, we use available CO and \HI\ observations.  
For the \blos , we use the catalog of \citet{Tahanietal2018}. We expand on these data below. 

\subsection{Galactic magnetic field}
We use the \citet[][hereafter referred to as JF12]{JanssonFarrar2012} Galactic magnetic field (GMF) model to estimate the direction of the initial magnetic fields. The JF12 model contains a two-dimensional (2D) thin-disk field component closely associated with the Galactic spiral arms, an azimuthal/toroidal halo field component, and an X-shaped vertical/out-of-plane field component. In addition to the regular field, the JF12 model includes an optional striated random field component. \citet{JanssonFarrar2012} constrained their random model parameters based on Faraday rotation measurements and polarized synchrotron radiation from the Wilkinson Microwave Anisotropy Probe (WMAP) seven-year release (WMAP7) synchrotron emission data.  

To estimate the GMF in the vicinity of the Perseus molecular cloud, we use the Hammurabi program~\citep{Waelkensetal2009Hammurabi}. The Hammurabi program\footnote{\url{http://sourceforge.net/projects/hammurabicode/}}  is a synchrotron modeling code, that has been used in several works~\citep[e.g.,][]{Plancketal2016Galactic, JanssonFarrar2012} to produce  simulated maps, which are then compared to data and used to constrain the model. These models are included with the code, and we can use the Hammurabi code to estimate the GMF vectors in Cartesian coordinates at a particular location in the Galaxy corresponding to the position of the molecular cloud.

\subsection{Velocity information}
We consider both the available CO and \HI\ velocities. The CO and \HI\  observations enable us to explore the line-of-sight velocities of the molecular and the atomic regions, respectively. 

\textbf{\HI\ data:} For \HI\ velocity information, we use the all-sky \HI\ database of the \HI\ $4\pi$ Survey \citep[HI4PI;][]{HI4PICollaboration2016} with an angular resolution of $16.2'$. HI4PI is based on the Effelsberg-Bonn \HI\ Survey \citep[EBHIS;][]{EBHIS_Kerpetal2011, EBHIS_Windeletal2016} using the Effelsberg 100-m telescope, and the Galactic All-Sky Survey \citep[GASS;][]{McClureGriffithsetal2009} made with the Parkes 64-m telescope. The HI4PI survey has a spectral resolution of $1.49$\,km\,s$^{-1}$, channel separation of $1.29$\,km\,s$^{-1}$, and velocity range of $|v_{LSR}| \leq 600$\,km\,s$^{-1}$ for the northern (EBHIS) and $|v_{LSR}| \leq 400$\,km\,s$^{-1}$ for the southern (GASS) parts in the Local Standard of Rest frame (LSR). The significant overlap between EBHIS and GASS ($-5^{\circ} \leq \delta \leq 0.5^{\circ}$) allowed for the accurate merging of the two surveys in creating the HI4PI survey. 

\textbf{CO data:} To approximate the motion of each cloud, we used the radial velocities from the \citet{Dameetal2001} carbon monoxide survey. This catalog is a survey of the $^{12}$CO J(1-0) spectral line of the Galaxy at 115\,GHz  with the pixel spacing of $7.5'$ and velocity resolution of $1.3$\,km\,s$^{-1}$ (obtained with the CfA $1.2$-m telescope and a similar telescope on Cerro Tololo in Chile).

\subsection{Line-of-sight magnetic field}
We use the \blos\ catalog of \citet[][]{Tahanietal2018}. 
They used Faraday rotation measurements to determine \blos\ in and around four filamentary molecular clouds.  
To find \blos, they used a simple approach based on relative measurements to estimate the amount of rotation measure induced by the molecular clouds versus that from the rest of the Galaxy. They then determined \blos\ using the rotation measure catalog of \citet{Tayloretal2009}, 
a chemical evolution code, and the \citet{Kainulainenetal2009} extinction maps~\cite[for more details, see][]{Tahanietal2018}. They found that the \blos\ direction in the Perseus molecular cloud reverses from one side of the cloud to the other, as shown in Figure~\ref{fig:PerseusBlosBpos}. 

\section{Main results: reconstructing the 3D magnetic field morphology of the Perseus cloud}
\label{sec:3Dmorphologies}

Recent studies~\citep{Doietal2021, Bialyetal2021} suggest that the Perseus molecular cloud  formed as the result of interaction with bubbles (i.e., the SCI model). \citet{Bialyetal2021} suggest that multiple supernovae have created a bubble resulting in formation of the Perseus molecular cloud. The presence of bubbles and their contribution to the formation and evolution of the Perseus cloud, along with the observed \blos\ reversal, as shown in Figure~\ref{fig:PerseusBlosBpos}, suggest a model for the arc-shaped magnetic field morphology based on shock interaction in this region. Using the initial magnetic field vectors in this region and the orientation of \blos\ reversal, we can reconstruct the complete 3D morphology of the arc-shaped magnetic field associated with the Perseus cloud.

\subsection{Galactic magnetic field in the region}
\label{GMF}

\begin{figure}[htbp]
\centering
\includegraphics[scale=0.56, trim={0cm 0cm 0cm 0cm},clip]{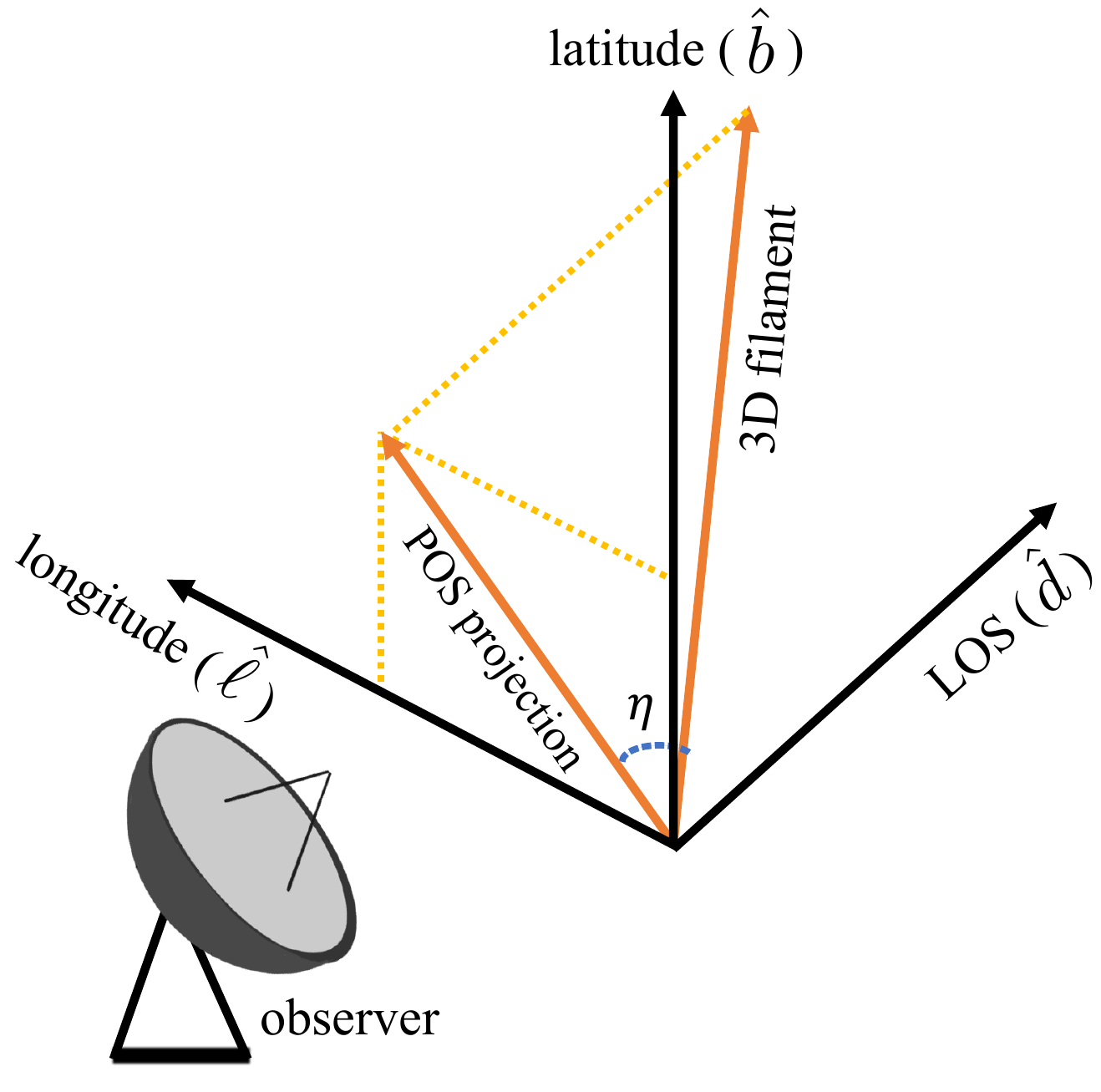}
\caption{The reference frame in our study. Longitude ($\hat{\ell}$) and latitude ($\hat{b}$) axes are pointing in the increasing direction of Galactic longitude and latitude (at the center location of the Perseus cloud in the plane of the sky), respectively. The line-of-sight ($\hat{d}$) axis points in the direction of increasing distance from us. The angle $\eta$ is the inclination angle of the filament with respect to the plane of the sky. } 
\label{fig:RefFrame}
\end{figure}

To determine the initial magnetic field direction, we use a GMF model based on JF12. This GMF model does not include the isotropic random field component, i.e, small-scale Gaussian random field~\citep[][]{Jaffeetal2010}, which 
arises from the zeroth-order simplification of the ISM turbulence and other environmental elements~\citep[][see their Figure~1]{Haverkorn2015, Jaffe2019}. We refer to this structure as the ``Coherent GMF'' model.  To best describe the GMF vectors and compare them with the \blos\ and \bperp\ observations, we use a frame of reference as shown in Figure~\ref{fig:RefFrame}, where the $\hat{\ell}$, $\hat{b}$, and $\hat{d}$ axes point in the increasing longitude, latitude, and distance directions (of the Perseus cloud's approximate center location), respectively. 

\begin{figure*}[htbp]
\centering
\includegraphics[scale=0.30, trim={0cm 0.7cm 0cm 1.5cm},clip]{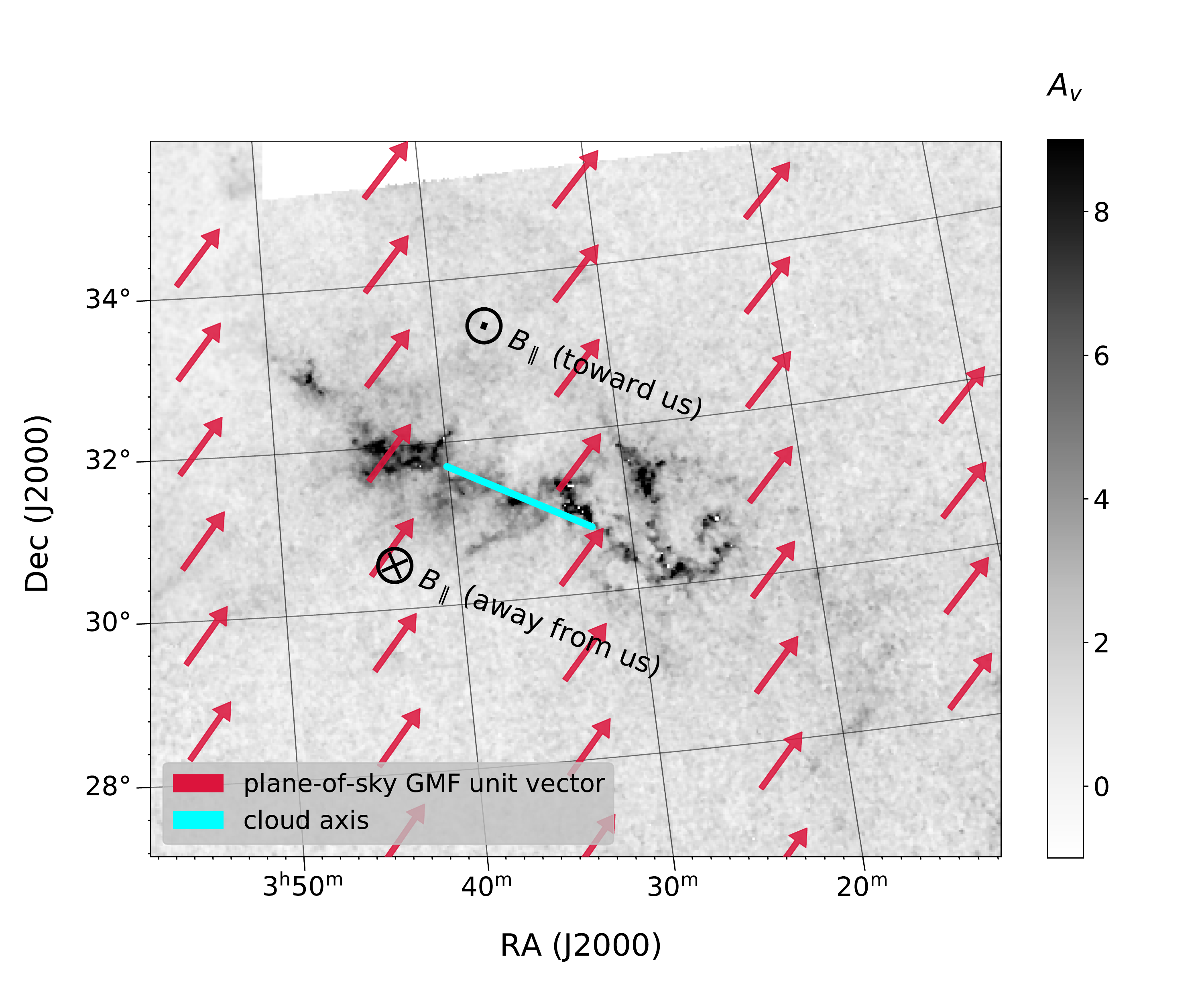}
\includegraphics[scale=0.30, trim={0cm 0cm 0cm 0cm},clip]{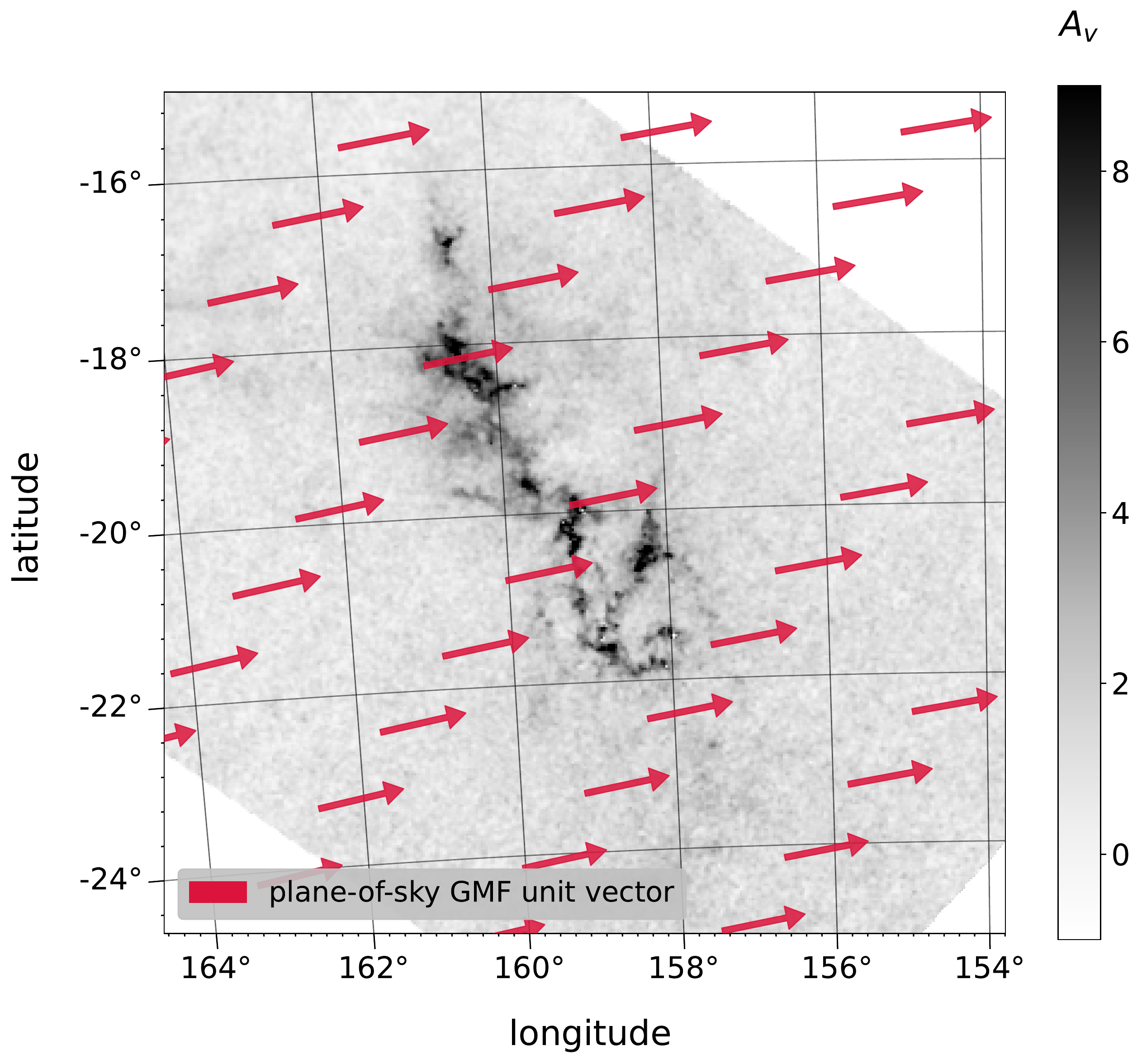}
\caption{Plane-of-sky GMF unit vectors over the Perseus molecular cloud showing the Coherent GMF model obtained from the Hammurabi code. The background gray-scale image shows the extinction map from~\citet{Kainulainenetal2009}. The red vectors show the GMF unit vectors in the plane of the sky. \textbf{Left panel:} The cyan line represents the approximate filament axis, with \blos\ directions reversing in the direction perpendicular to this axis. \textbf{Right panel:} For ease of comparison with Figure~\ref{fig:3DBPerseus}, the Perseus cloud and GMF vectors are shown in Galactic coordinates.} 
\label{fig:GMFPers}
\end{figure*} 

We delineate a volume around the Perseus molecular cloud to model the GMF, $\sim200$\,pc deep along the line of sight (with $\sim$100\,pc in front of and behind the 294\,pc average cloud distance), spanning $\sim 11^{\circ} \times 11^{\circ} $ (104 pc$^{2}$) in the plane of the sky. We set a resolution of one GMF vector per $2\rm{\,pc}\times 2\rm{\,pc}\times 2\rm{\,pc} $. Figure~\ref{fig:GMFPers} illustrates the Plane-of-sky GMF unit vectors (red arrows) at the location of the Perseus cloud for this Coherent GMF model, overlaid on the extinction map of this cloud. These also match previous studies and their modeled GMF vectors at this location~\citep[e.g.,][see their Figure 6]{VanEcketal2011}.  
We find that the Coherent GMF lies mostly perpendicular to the main axis of Perseus. We estimate the unit vector showing the direction of the Coherent GMF in our frame of reference to be  $ -0.99 \hat{\ell} + 0.15 \hat{b} -0.01 \hat{d}$. 

\subsection{Complete 3D magnetic field morphology of the Perseus cloud}
Knowing a high likelihood for the presence of an arc-shaped magnetic field morphology in this region, we reconstruct the complete 3D magnetic field morphology of this cloud. For this purpose, we use the \blos\ observations and the Coherent Galactic magnetic field components in this region. 

Ideal magnetohydrodynamic (MHD) simulations ~\cite[e.g.,][]{LiKlein2019} show that strong enough magnetic fields (with Alfv\'en mach number $\mathcal{M}_A \simeq 1$) retain a memory of the initial large-scale magnetic field lines.  
If the initial magnetic fields are weak ($\mathcal{M}_A \simeq 10$), the magnetic fields will get completely distorted, not following any particular large-scale morphology. 

Both the \blos\  \citep{Tahanietal2018} and the \bperp\ \citep{PlanckXXXV} observations show that the large-scale magnetic fields associated with the Perseus cloud are coherent, as illustrated in Figure~\ref{fig:PerseusBlosBpos}. The \bperp\ lines are mostly perpendicular to the cloud's axis and the \blos\ reverses direction from one side of the cloud's axis to the other. The \bperp\ orientation for the Perseus cloud, as studied by the histograms of relative orientation in \citet{PlanckXXXV} and \citet{Soler2019}, is mostly perpendicular to the cloud's axis (when projected onto the plane of the sky) and overall in the same orientation as our modeled Coherent GMF lines. These large-scale coherent magnetic fields  indicate that the initial field lines have become more ordered and stronger during the cloud's formation and evolution. In other words, these magnetic fields retained a memory of the initial field lines and have not become completely distorted. We discuss this in a more quantitative manner in Section~\ref{discussion}. 

\begin{figure*}[htbp]
\centering
\includegraphics[scale=0.53, trim={2cm 0cm 1cm 3cm},clip]{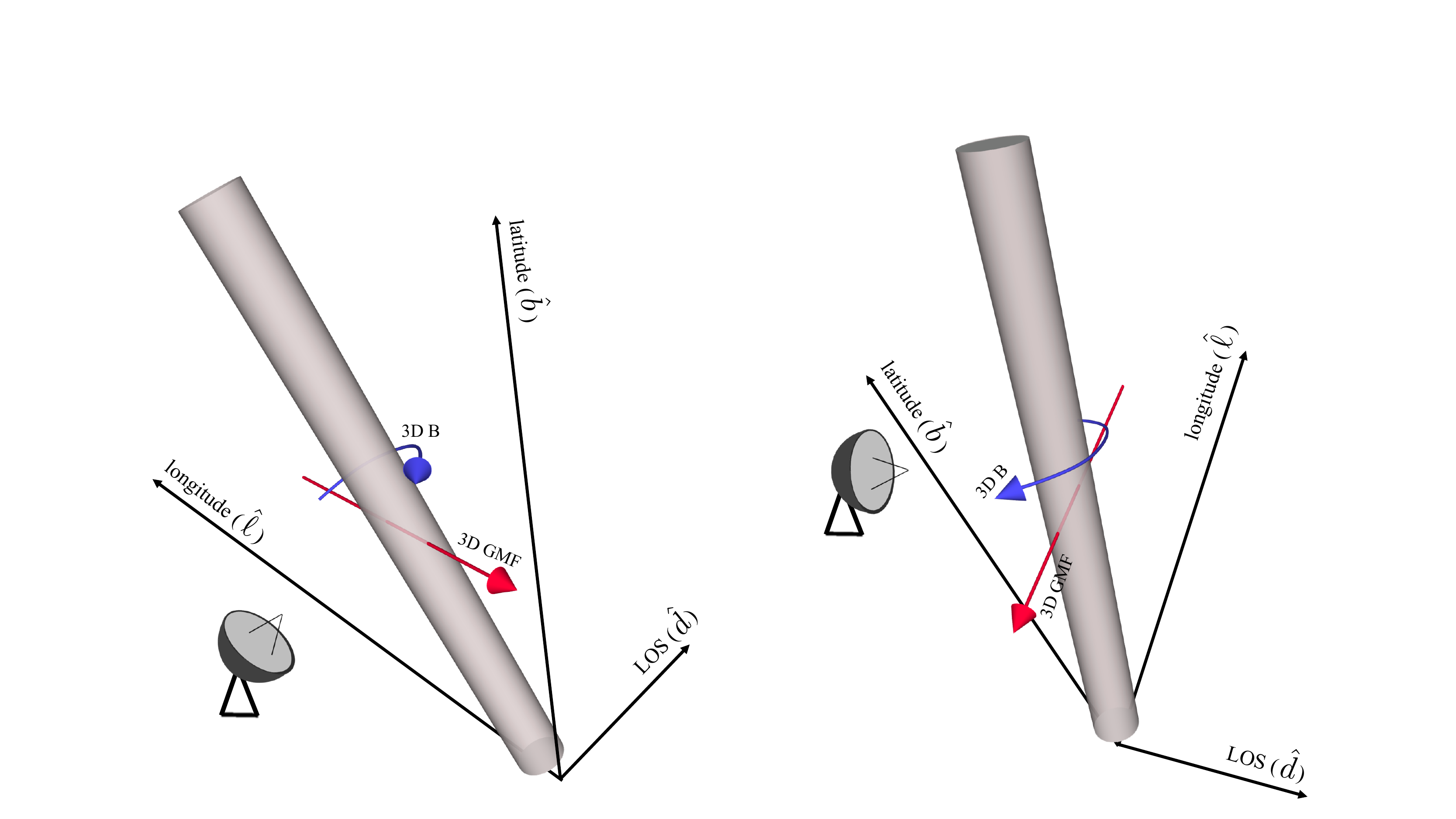}
\caption{The 3D magnetic field morphology of the Perseus cloud. The gray cylinder, the red vector, and the blue arrow illustrate the filamentary cloud, the estimated Galactic magnetic field, and the reconstructed 3D magnetic field morphology associated with the cloud, respectively. The 3D magnetic field morphology is concave from the observer's point of view. The left and right panels show different viewing angles.} 
\label{fig:3DBPerseus}
\end{figure*}

The direction of the GMF component, as shown in Figure~\ref{fig:GMFPers}, is mostly perpendicular to the Perseus cloud filament axis and pointing from southeast side of the cloud to northwest in the equatorial coordinate system. Since the cloud preserves a memory of the large-scale GMF, this GMF direction, along with the \blos\ directions, indicate that the arc-shaped field morphology must be concave from the observer's point of view, with its plane-of-sky component pointing in the $-\hat{\ell}$ direction.  
Figure~\ref{fig:3DBPerseus} shows a cartoon diagram of the complete 3D morphology of the magnetic fields in the Perseus cloud\footnote{The .obj files are available at~{\url{https://github.com/MehrnooshTahani/Perseus3DMagneticFields}}}. 

This 3D magnetic field morphology results in  magnetic field lines projected on the plane of the sky that are consistent with the Planck \bperp\ observations. We note that the distances along the Perseus cloud~\citep{Zuckeretal2018, Zuckeretal2020, Zuckeretal2021} suggest a relatively small inclination angle for the Perseus cloud. 
While we consider the cloud to be nearly parallel to the plane of the sky, larger inclination angles have no significant effect on the reconstructed 3D magnetic field morphology.

\section{Comparison of the result with cloud-formation model predictions}
\label{sec:formationScenario}
\begin{figure*}[htbp]
\centering
\includegraphics[scale=0.6, trim={0cm 0cm 0cm 0cm},clip]{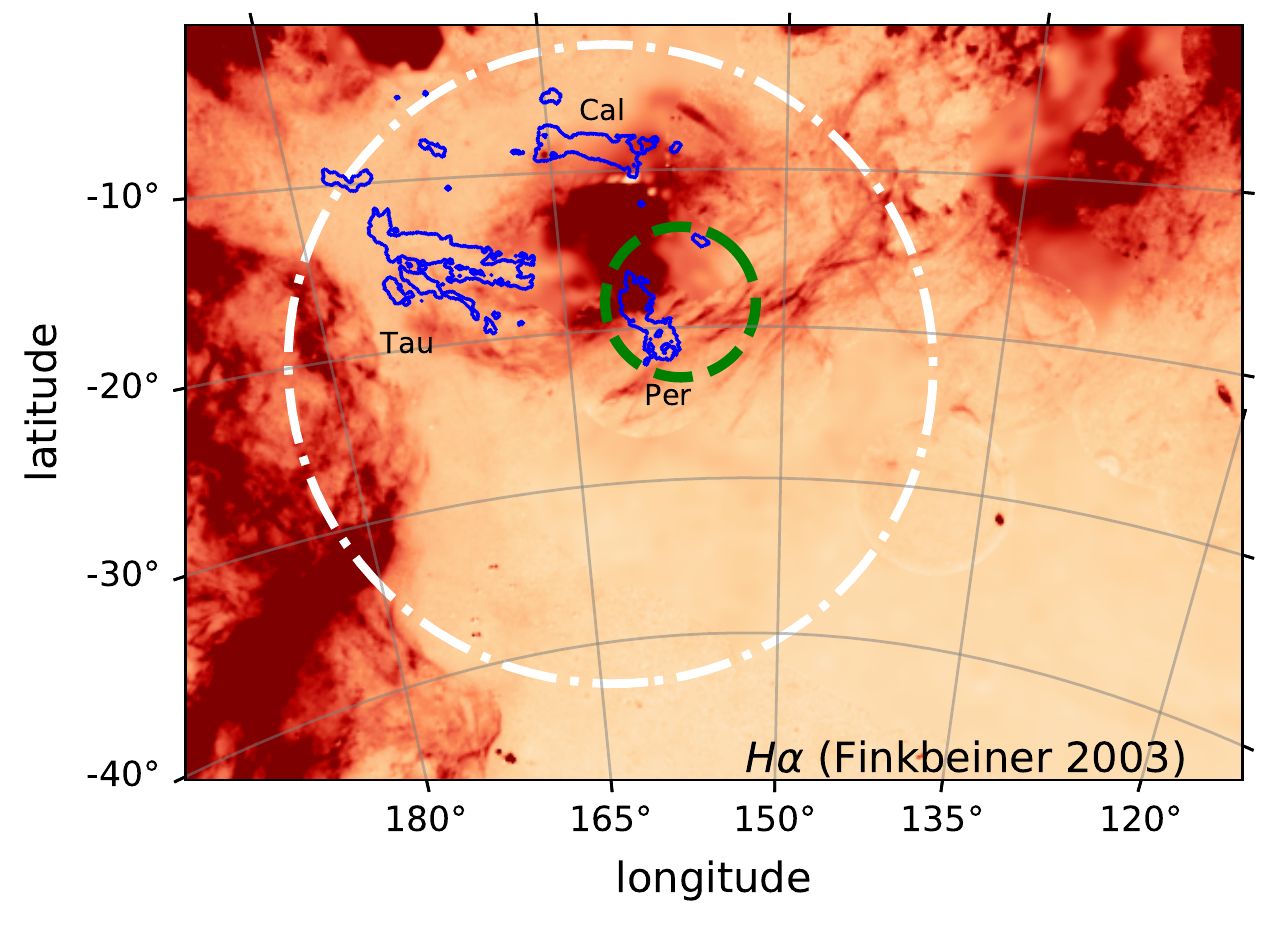}
\includegraphics[scale=0.6, trim={0cm 0cm 0cm 0cm},clip]{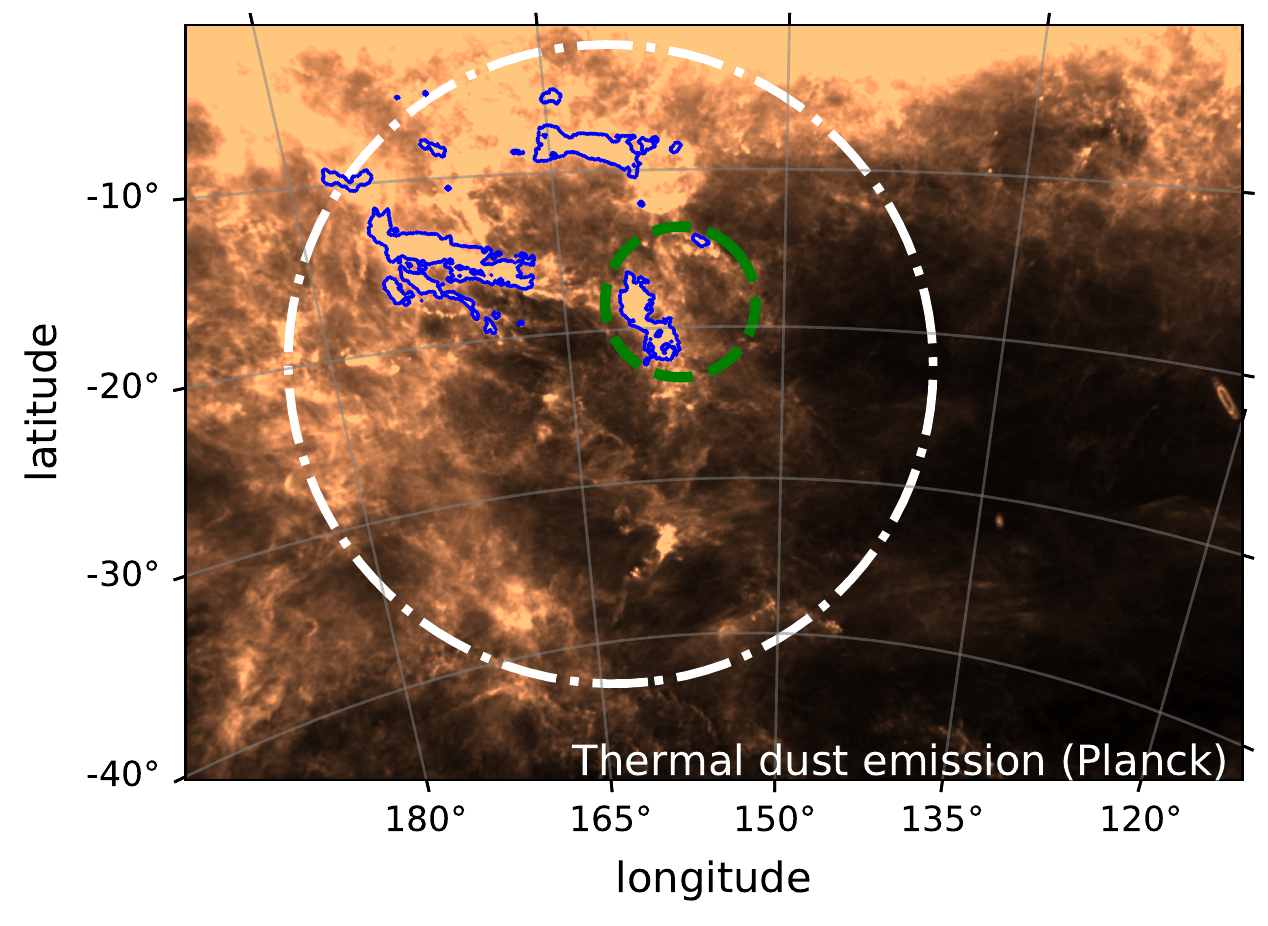}\\
\includegraphics[scale=0.6, trim={0cm 0cm 0cm 0cm},clip]{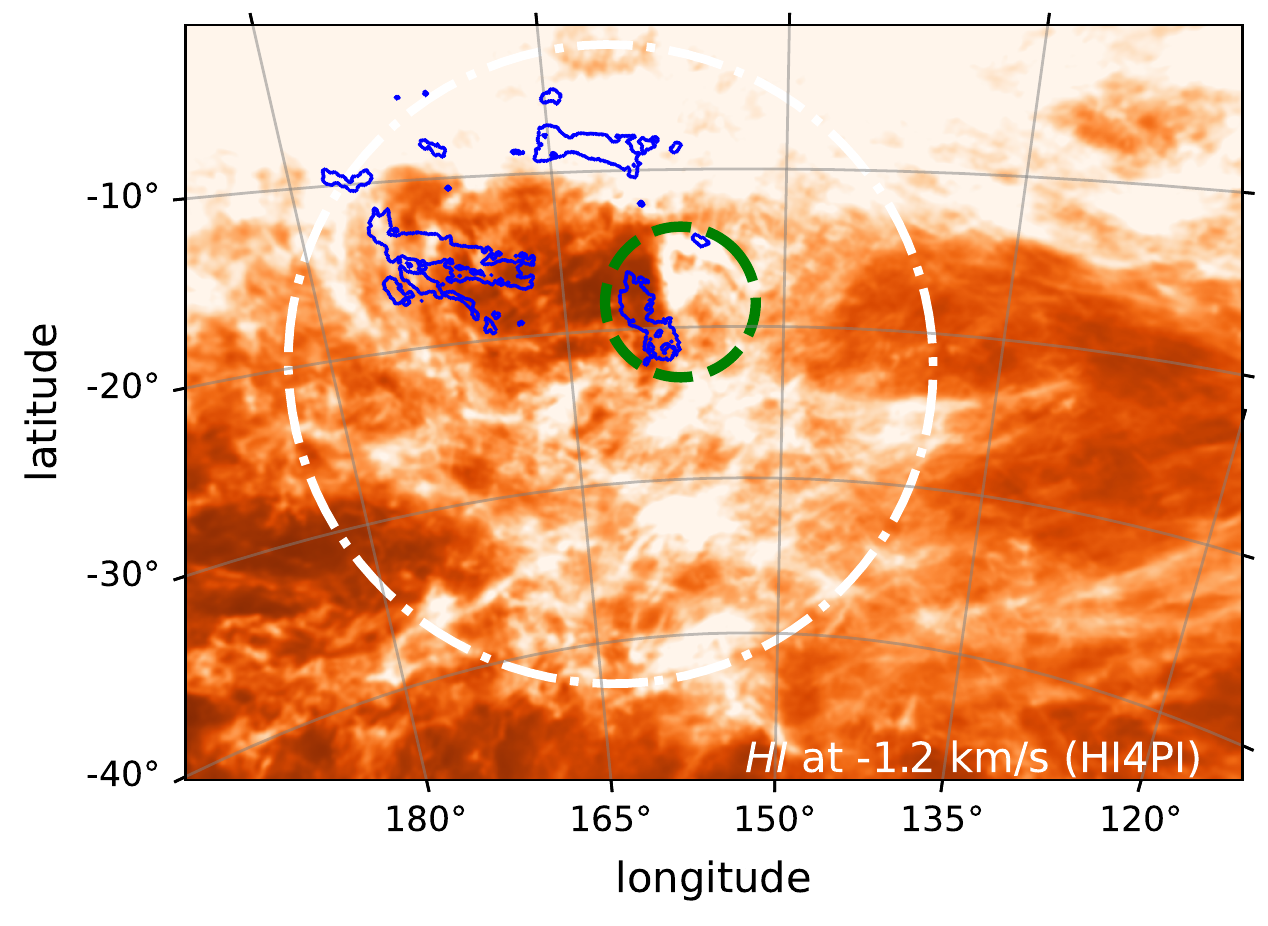}
\includegraphics[scale=0.6, trim={0cm 0cm 0cm 0cm},clip]{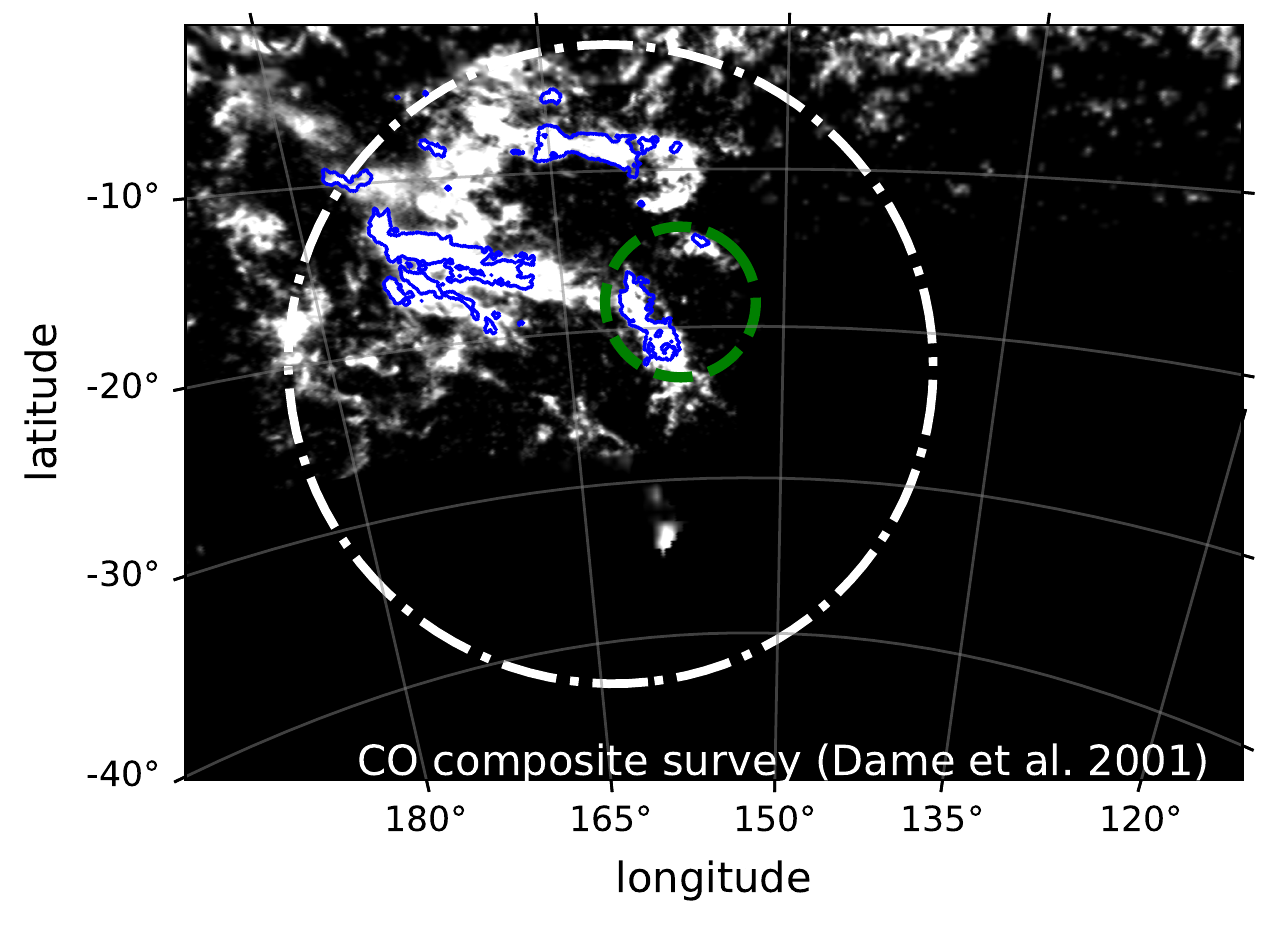}\\
\caption{Multi-wavelength observations of the Perseus cloud and associated bubbles. Herschel contours of the Perseus, Taurus, and California clouds are shown in blue (denoted in the top-left panel as Per, Tar, and Cal, respectively). The Per-Tau shell \citep{Bialyetal2021} is depicted as a white dot-dashed circle and the Per2 shell is approximated as the green dashed circle. \textbf{Top-left panel:} $H\alpha$ observations \citep{Finkbeiner2003}. \textbf{Top-right panel:} Dust observations from the Planck Space Observatory. \textbf{Lower-left panel:} \HI\ observations at a velocity of $-1.2$ km/s from HI4PI. \textbf{Lower-right panel:} Composite CO survey of \cite{Dameetal2001}. } 
\label{fig:MultiWaveObsPerseus}
\end{figure*}

\begin{figure}[htbp]
\centering
\includegraphics[scale=0.15, trim={12cm 10cm 20cm 20cm},clip]{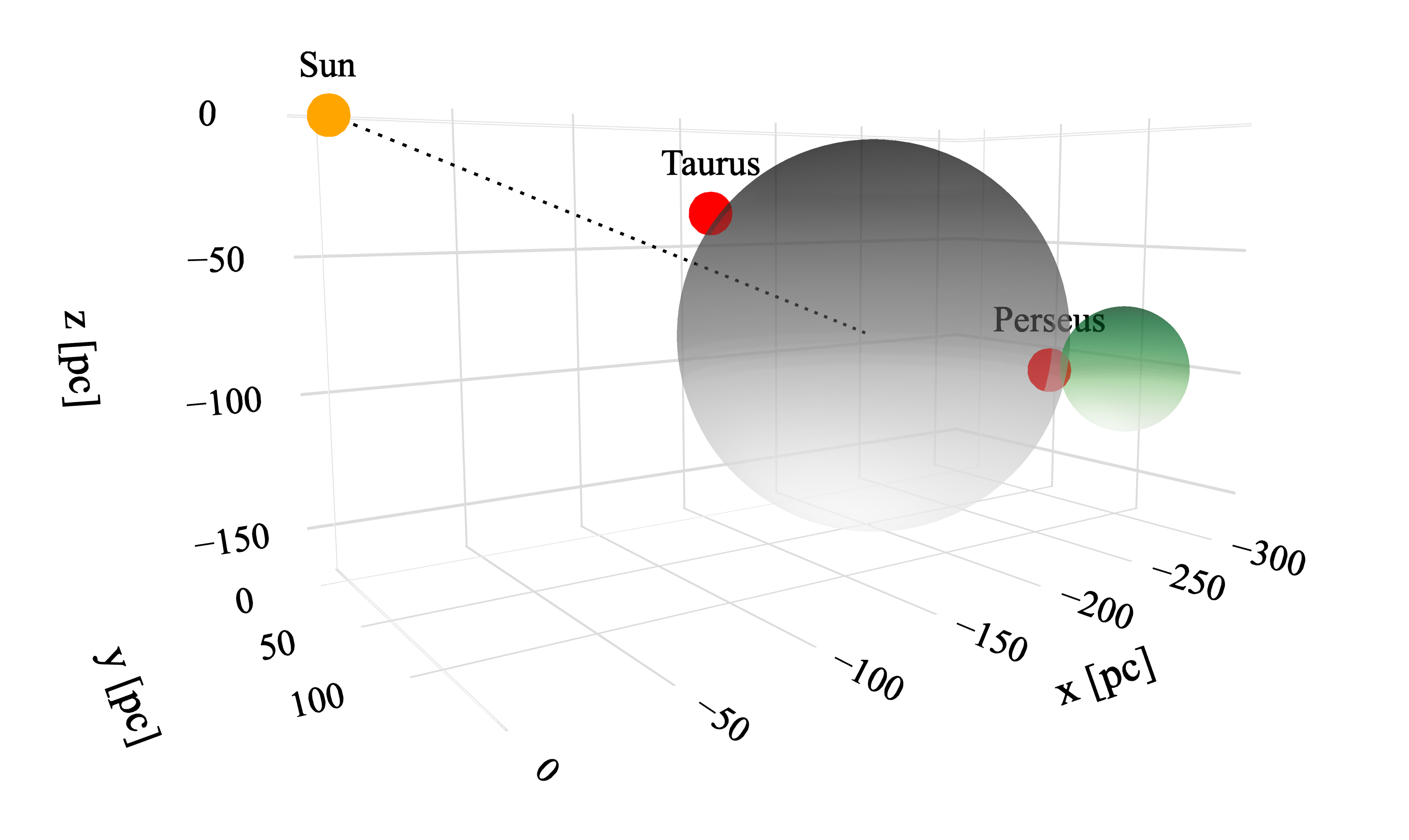}
\caption{3D approximation of the region surrounding the Perseus and Taurus molecular clouds.  The approximate locations of the Perseus and Taurus molecular clouds are shown as red circles, and the Per-Tau shell \citep{Bialyetal2021} is shown as a gray sphere. The Sun is depicted as a yellow circle and the line-of-sight path connecting the Sun to the center of the Per-Tau shell is shown as a dashed line. The Per2 shell-like structure is approximated as a sphere located behind the Perseus cloud, shown in green. The size of the Taurus and Perseus clouds and the Sun are not depicted to scale. } 
\label{fig:3DShells}
\end{figure}

\citet[][see their Figures 10 and 11]{Doietal2021} identify a dust cavity in this region, where the Perseus and Taurus clouds are located on the far- and near-sides of its shell, respectively. \citet[][see their Figure~2]{Bialyetal2021} provide 3D modeling of this bubble and refer to it as Per-Tau. \citet{Bialyetal2021} propose
that this bubble has swept up the interstellar medium and has ultimately led to the formation of the Perseus and Taurus molecular clouds. We illustrate this  bubble projected on the plane of the sky as a white  dash-dotted circle in Figure~\ref{fig:MultiWaveObsPerseus}, and  as a 3D gray sphere in  Figure~\ref{fig:3DShells}.

\citet{Bialyetal2021} suggest that this expanding shell has swept-up the interstellar medium and has ultimately led to the formation of the Perseus and Taurus molecular clouds. They also found that multiple supernovae have likely contributed to the expansion of this shell and explained the formation of the Perseus cloud through the multi-compression SCI model of  \citet{Inutsukaetal2015}. In this scenario, an initial  bubble expansion can accumulate the ISM on a bubble's shell creating a relatively dense cloud. Subsequently, recurrent compressions from other (or the same) bubbles can result in a dense filamentary molecular cloud. 

Additionally, \citet{Ponetal2016} observed an excess emission of the CO J(6-5) transition in the Perseus molecular cloud, indicating the presence of low-velocity 
shocks (and their turbulence dissipation).  
\citet[][]{Ponetal2014} found similar CO J(6-5) emission in Barnard~1 \citep[B1; ][]{BachillerCernicharo1984} region within the Perseus cloud. The confirmed presence of shocks in the Perseus cloud provides additional evidence for the formation and evolution of the Perseus cloud as predicted by the SCI model.

Using the  SCI model, if we estimate the initial magnetic field direction as well as the velocity of the dense cloud (CO) and its \HI\ surroundings,  we will be able to predict the \blos\ direction associated with the cloud and its 3D magnetic field morphology (as shown in Figure~\ref{fig:SCIModel}). We can compare these predictions with the \blos\ observations of~\citet{Tahanietal2018} and our reconstructed magnetic fields. To this end, we use the Coherent GMF direction to estimate the initial magnetic field direction in the region, and assume that the cloud retains a memory of the most recent shock, interaction, or compression. To properly define the velocities, we consider the average velocity of the pre-existing cloud, \vec{v}, in the co-moving frame of the mean shock front. In general, the velocity is not aligned with the mean magnetic field, so we can assume that the field lines have a perpendicular component to the propagation direction, 
as illustrated in Figure~\ref{fig:SCIModel}. 

This interaction bends the magnetic field lines such that if \vec{v} (velocity of the dense cloud in the \HI\ co-moving frame) is pointing toward [away from] us, a convex [concave] bend in the field lines will be created from the observer's point of view. Additionally, the direction of this arc-shaped magnetic field, causing a \blos\ reversal, can be predicted. For example, if the initial magnetic field is pointing from side B to side A in Figure~\ref{fig:SCIModel}, the concave pattern will result in magnetic field lines pointing toward the observer on side A and away from the observer on side B, resulting in a \blos\ reversal across the filament. To compare the magnetic field observations with the cloud-formation model predictions, we investigate the velocities of the cloud. In this comparison, we assume that while the velocities prior to the interaction (between the cloud and the shock front) have largely dissipated, the current-state direction of the cloud's velocity in the \HI\ frame retains a memory of the most recent interaction.

\subsection{Velocities}
\label{velocities}

We first consider the Galactic rotation velocities in this region to paint a comprehensive picture. On the scales of a giant molecular cloud and the Galaxy, we can assume that the conditions for ideal MHD hold true~\citep{HennebelleInutsuka2019}   
and the field lines are frozen in the 
interstellar medium (the filamentary structure and its surroundings). 
We use the model of~\citet{Clemens1985_GalRot} with the IAU standard values of the solar distance from the Galactic center 
($8.5$\,kpc) and orbital velocity (220\,km\,s$^{-1}$), we find a Galactic rotation velocity (LSR) of $\sim -2$\,km\,s$^{-1}$ for the Perseus cloud (with the average longitude, $l$, of $159.5^{\circ}$ and the average distance, $d$, of 294\,pc).    
Although recent studies~\citep[e.g.,][]{Russeil2017_GalRot, Krelowskietal2018_GalRot} have revisited the standard Galactic rotation models of \citet{Clemens1985_GalRot} and \citet{BrandBlitz1993_GalRot}, the \citet{Clemens1985_GalRot} model is sufficient for our purposes.

\begin{figure*}[htbp]
\centering
\hspace{-0.5cm}
\includegraphics[scale=0.42, trim={0cm 0cm 0cm 0cm},clip]{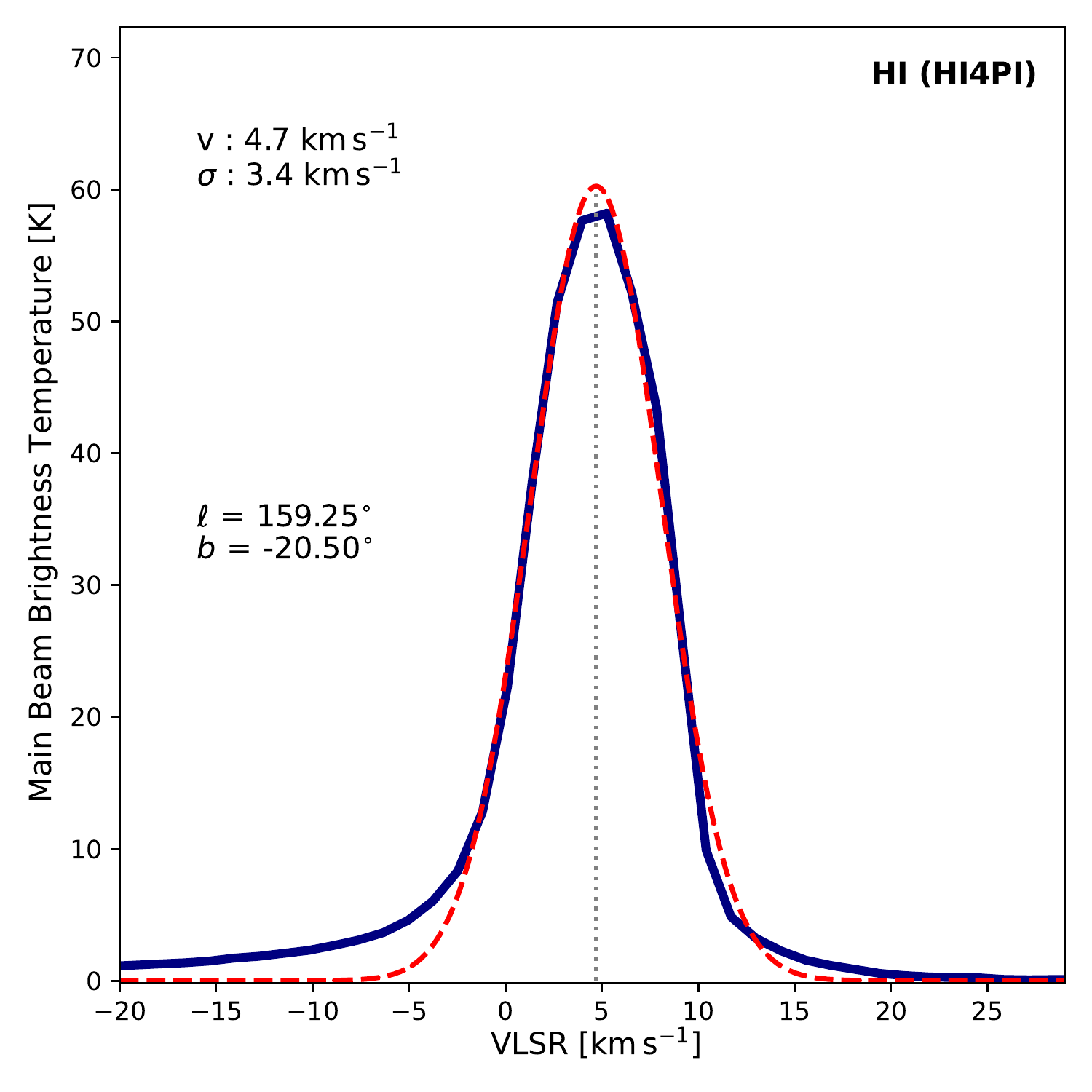}
\hspace{1cm}
\includegraphics[scale=0.42, trim={0cm 0cm 0cm 0cm},clip]{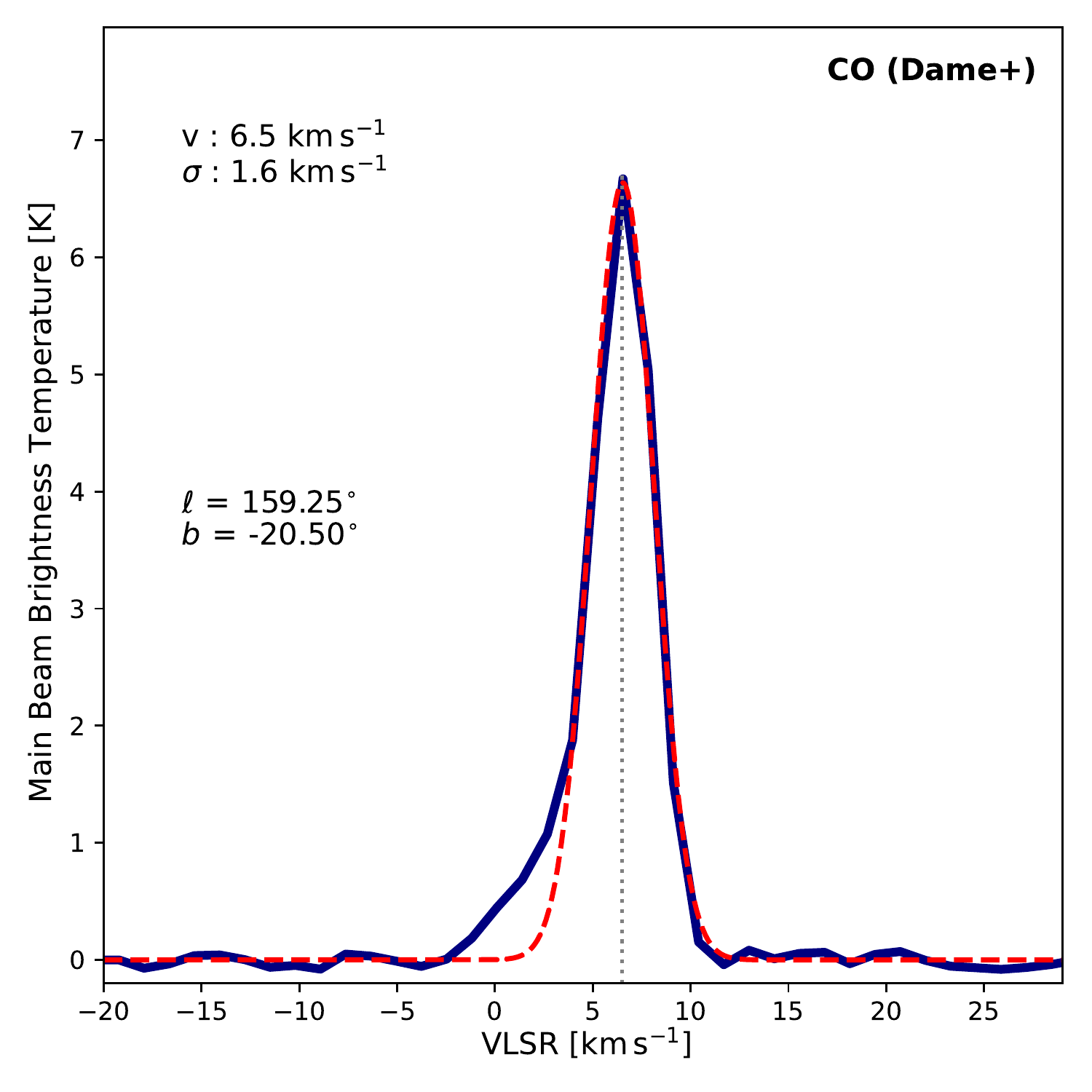}
\caption{\HI\ and CO spectra for the Perseus molecular cloud for one coordinate. The blue line shows the \HI\ spectral line obtained from the HI4PI survey in the left panel and the observed $^{12}$CO J(1-0) line obtained by \citet{Dameetal2001} in the right panel. The red line shows a Gaussian fit to each spectrum. The $x$ axis shows the LSR velocity and the $y$ axis shows the main-beam brightness temperature. In this work, we are interested in the peak emission by CO and \HI\ averaged along the entire cloud.} 
\label{fig:VelocityProfile_Perseus}
\end{figure*}

To estimate the average \HI\ velocity of the cloud, we study the \HI\ spectrum at various points on the cloud. 
We pick locations for which the \HI\ spectrum shows a single \HI\ peak; i.e., does not exhibit multiple distinguishable peaks, absorption, or significant self-absorption patterns. The \HI\ emission spectrum of these points can be mainly described by a single Gaussian fit.  
To estimate the average CO velocity, we smooth the CO data to the same spatial full width at half maximum as the \HI\ data and re-grid the CO data cube to match the \HI\ cube. With the smoothed and re-gridded CO data, we follow the same approach as explained for \HI\ velocities to obtain the average CO velocity of the cloud.   
Figure~\ref{fig:VelocityProfile_Perseus} shows an example of these spectra (for \HI\ and CO) for only one sky coordinate in the Perseus cloud. In this figure, the blue line shows the \HI\ line obtained from the HI4PI survey in the left panel, and the observed $^{12}$CO J(1-0) line obtained from \citet{Dameetal2001} in the right panel. The red line shows a Gaussian fit to each spectrum. The $x$ axis shows the LSR velocity and the $y$ axis shows the main-beam brightness temperature. 

To find the molecular cloud velocities (CO) in the co-moving frame of the \HI\ gas we use:
\begin{equation}
\varv_{\text{CO - H\protect\scaleto{$I$}{1.2ex}}} = \varv_{\text{CO, LSR}} - \varv_{\text{H\protect\scaleto{$I$}{1.2ex}, LSR}},
\label{eq:CoFrame}
\end{equation} 
where \vcoHi\ is the cloud CO velocity in the co-moving \HI\ frame, \vco\ is the CO velocity in the LSR frame and \vhi\ is the \HI\ velocity of the region in the LSR frame. 

In this work, we are interested in the peak emission by CO and \HI\ averaged throughout the entire cloud. To estimate the CO velocities in the co-moving \HI\ frame, we adopt three different approaches: 1) {\textbf{Averaging CO and \HI\ velocities separately:}} In this approach, we find both the \HI\ and CO velocities of a number of points across the cloud, determine the average CO and  \HI\ velocities, separately, and finally use equation~\ref{eq:CoFrame} to find \vcoHi\ of the cloud. We find the standard deviation of the averaged velocity centroids ($\sigma_{\text{total}}$) to estimate the uncertainty in the mean CO and \HI\ velocities. 2) {\textbf{Finding correlation coefficient:}} The \HI\ velocities associated with the cloud are more ambiguous than CO due to more line-of-sight confusion and blending effects. To confirm the \HI\ velocity of the cloud, we find the correlation coefficients between \HI\ beam brightness temperature at each velocity slice of the HI4PI data cube and \Av\ values. This approach allows us to find an LSR velocity that corresponds to locations with higher extinction values and is directly linked to the molecular cloud. By studying the \HI\ spectra once separately on the cloud as explained earlier  
and once using their correlation with \Av , we ensure that the \HI\ emissions are linked to the cloud of study. 3) {\textbf{Obtaining \vcoHi\ point by point:}}  
We find \vcoHi\ for each point (with simple \HI\ and CO spectra) on the cloud separately, and determine the average and the standard deviation of all these \vcoHi\ values. This approach accounts for any velocity variations and gradients across the cloud. 

\begin{figure*}[htbp]
\centering
\includegraphics[scale=0.30, trim={0cm 1.5cm 0cm 1cm},clip]{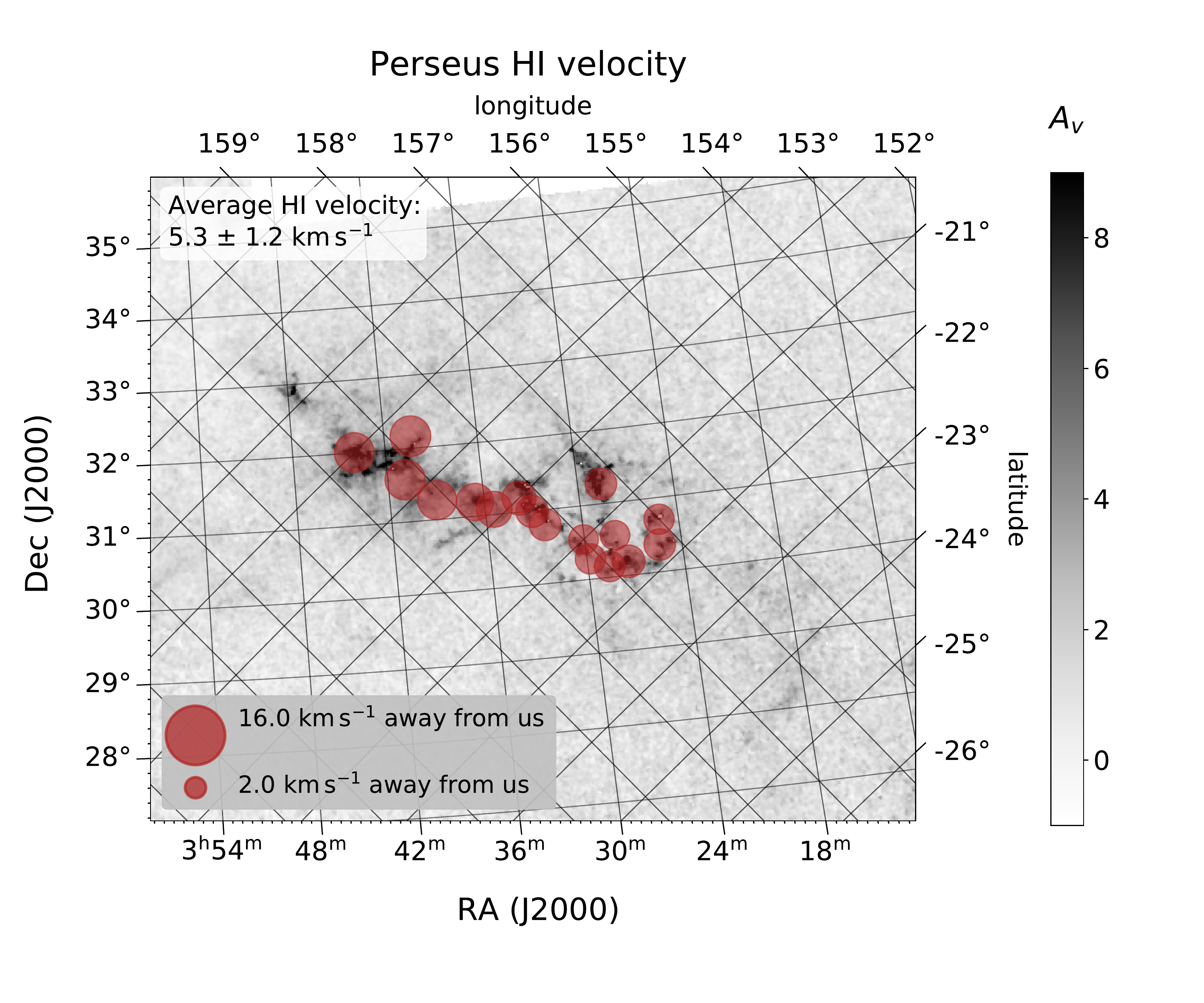}
\includegraphics[scale=0.30, trim={0cm 1.5cm 0cm 1cm},clip]{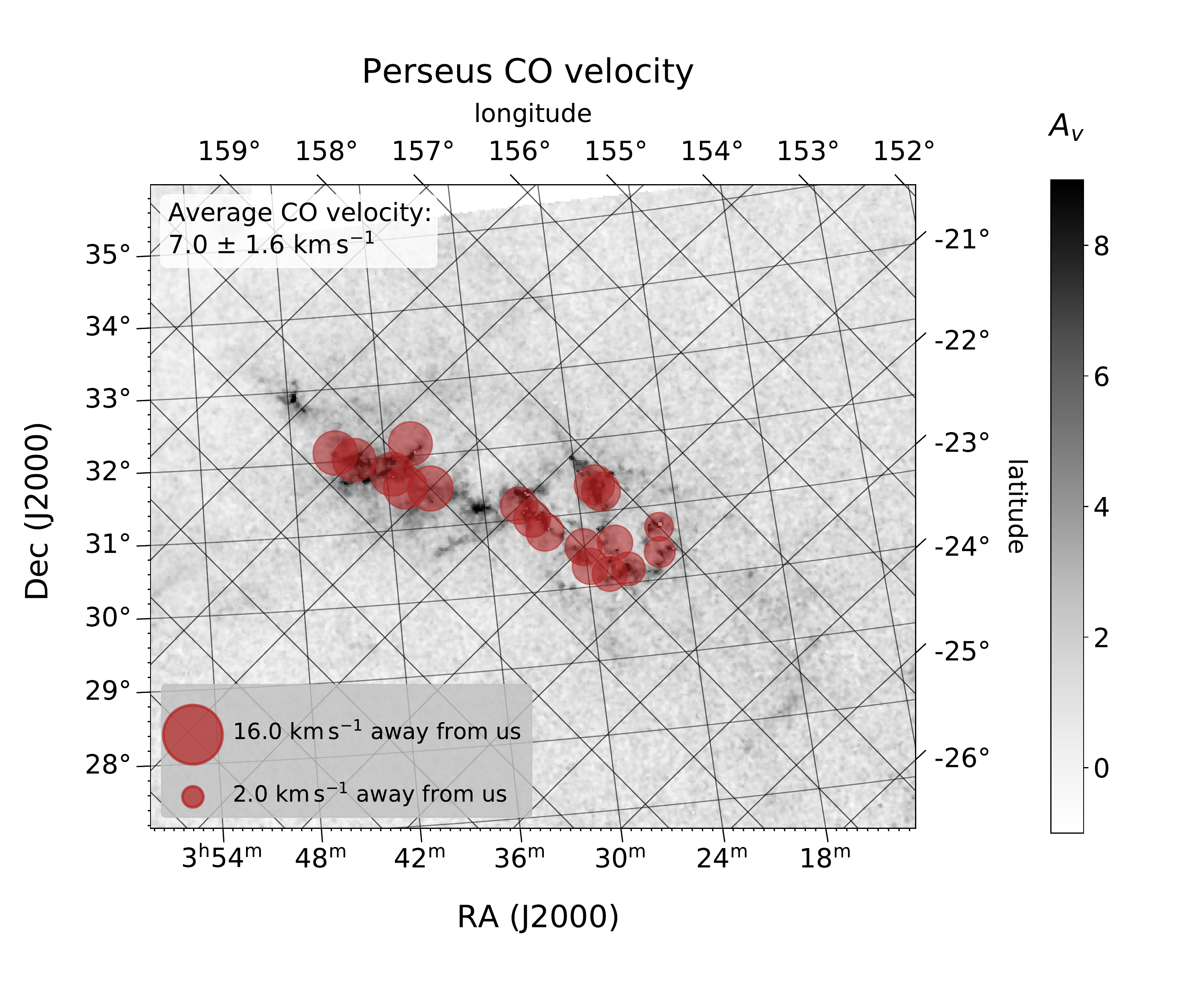}
\caption{Perseus molecular cloud \HI\ and CO velocities (LSR) from \citet{HI4PICollaboration2016} and \citet{Dameetal2001}, respectively. The red circles show velocities pointing away from us. This figure shows our first approach for determining the \vcoHi\ direction, as described in Section~\ref{velocities}. The size of the circles shows the magnitude of the velocities. The background gray-scale image shows the extinction map in \Av\ magnitude as obtained by \citet{Kainulainenetal2009}. \textbf{Left panel:} The circles show the \HI\ velocities. \textbf{Right panel:} The circles show the CO velocities.} 
\label{fig:HICOVelocity_Perseus}
\end{figure*}

Figure~\ref{fig:HICOVelocity_Perseus} shows the \HI\ (left panel) and the CO (right panel) velocity locations on the Perseus cloud, 
which yield an average LSR velocity of $5.3 \pm 1.2$\,km\,s$^{-1}$ and $7.0 \pm 1.6$\,km\,s$^{-1}$, respectively as explained in the first approach. In this figure, the red circles show LSR velocities away from us and the size of the circles shows their magnitude.  
The velocities also agree with different studies using \HI , $^{12}$CO and $^{13}$CO tracers, at similar locations~\citep{ImaraBlitz2011, Arceetal2011, Zuckeretal2018, Remyetal2018, Nguyenetal2019}. We note that the standard deviation values both in CO and \HI\ are largely associated with the velocity gradients along the main axis of the Perseus cloud (which is addressed in the third approach). 

\begin{figure}[htbp]
\centering
\includegraphics[scale=0.40, trim={0cm 0cm 0cm 0cm},clip]{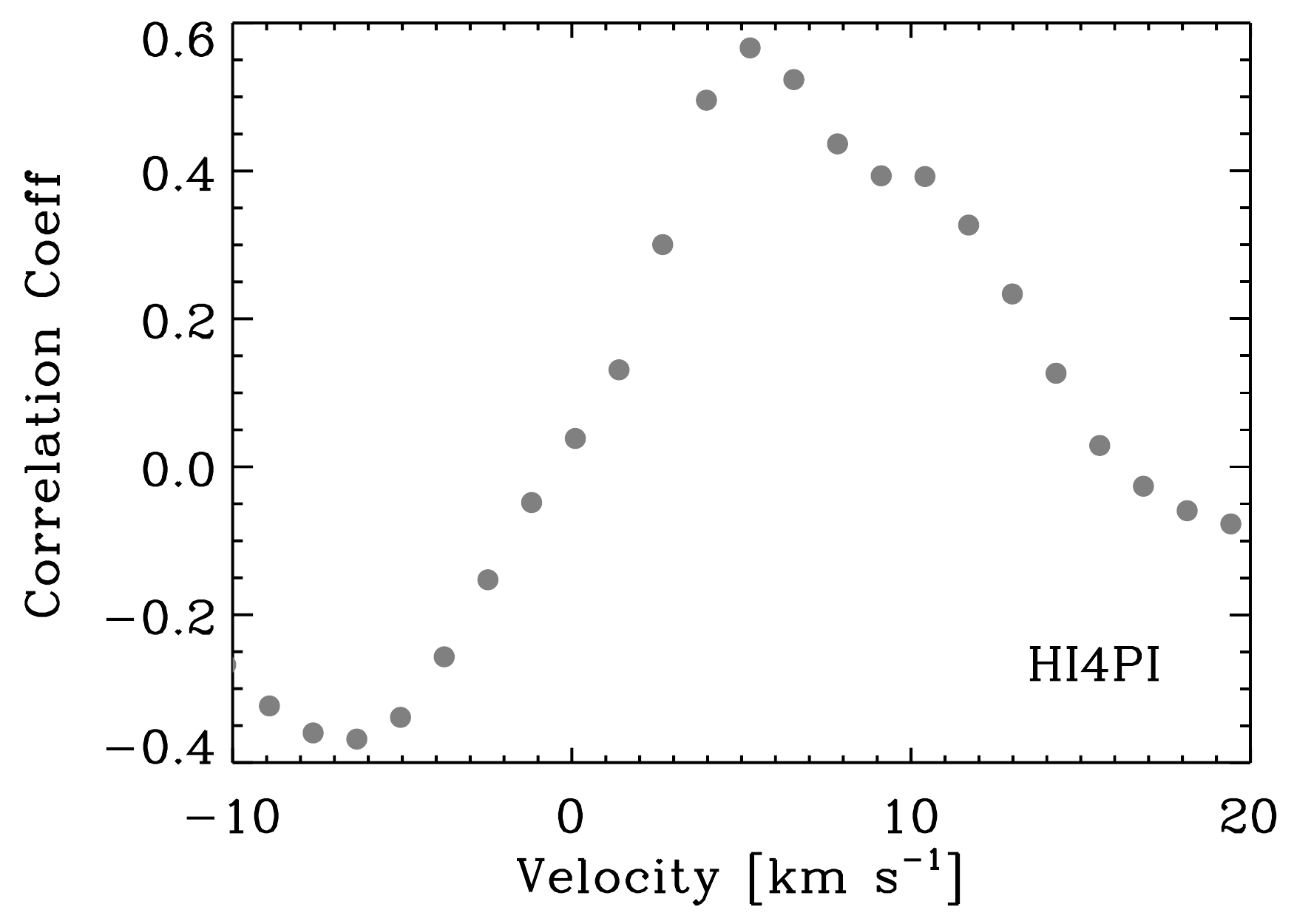}
\caption{Correlation coefficients between \HI\ and \Av\ in the Perseus cloud. To confirm the \HI\ velocities, we find the correlation coefficients between the \HI\ beam brightness temperature and \Av\ values, at each velocity slice. The $y$ axis shows the correlation coefficient and the $x$ axis shows the LSR velocity. This figure illustrates our second approach for determining the \vcoHi\ direction as described in Section~\ref{velocities}. } 
\label{fig:HIAVCorr_Perseus}
\end{figure}

To confirm that this \HI\ velocity is associated with the Perseus molecular cloud, as described in the second approach, we find the correlation coefficients between \HI\ beam brightness temperature in each velocity slice and the \Av\ values. We find a peak in the correlation coefficients around an LSR velocity of $5$\,km\,s$^{-1}$, confirming that the average \HI\ velocity that we found in approach~1 is indeed associated with the Perseus cloud. Figure~\ref{fig:HIAVCorr_Perseus} shows the obtained correlation coefficients between \HI\ and \Av\ with a peak around $5$\,km\,s$^{-1}$ (LSR). We note that the \HI\ spectra for the Perseus cloud are relatively easier to analyze compared to other regions in the Galaxy due to its high latitude ($b \sim -20^{\circ}$). Using these approaches, we find that the average \HI\ velocity associated with the Perseus cloud is smaller than the CO velocity but both have the same direction. 

\begin{figure}[htbp]
\centering
\includegraphics[scale=0.46, trim={0cm 0cm 0cm 0cm},clip]{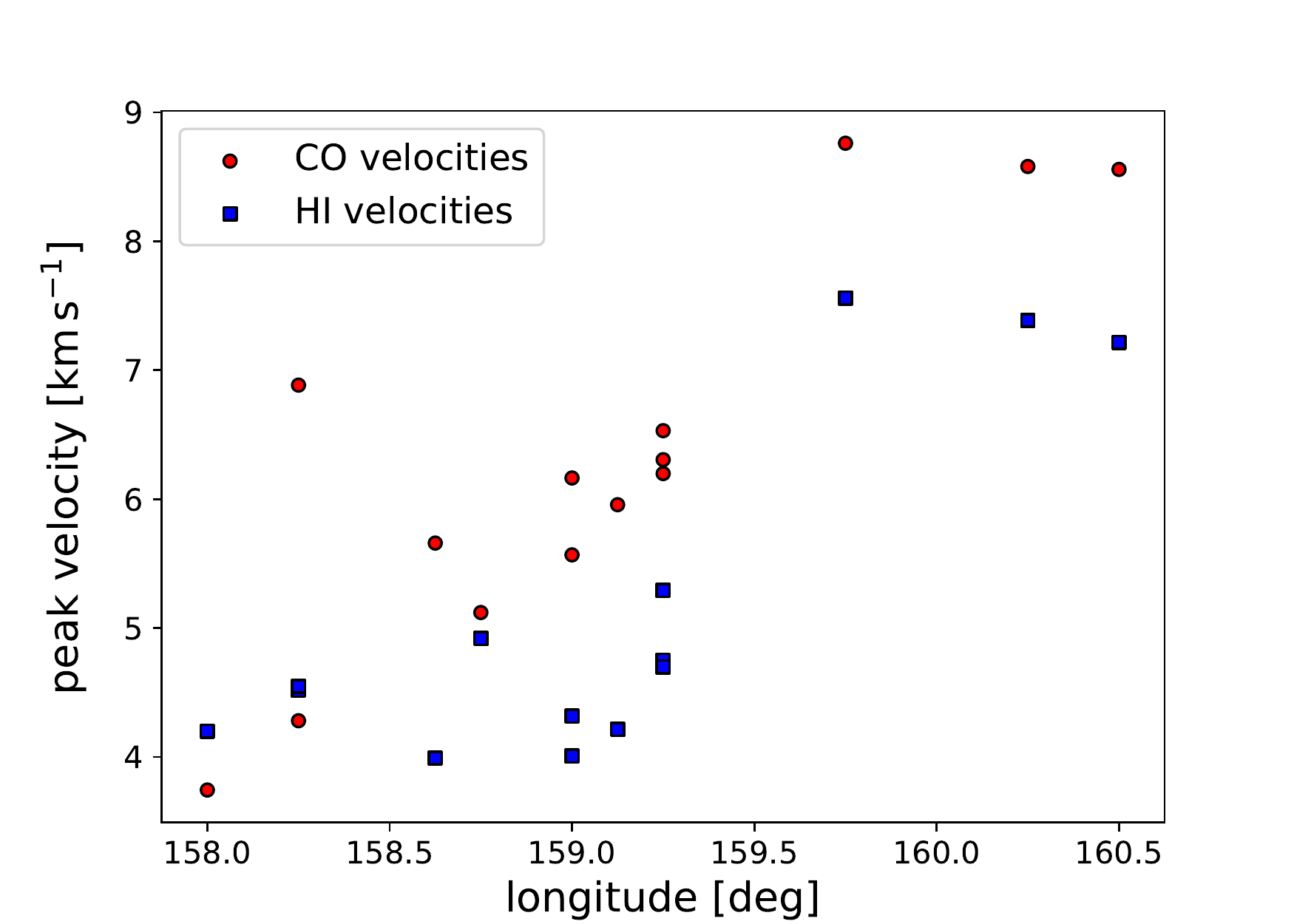}
\caption{CO and \HI\ velocities at peak emission in the Perseus cloud. The red circles and blue squares show the velocities at peak emission for CO and \HI , respectively. The $y$ axis shows the LSR velocity at peak emission and the $x$ axis shows the longitudes of points. This figure depicts our third approach for determining the \vcoHi\ direction as described in Section~\ref{velocities}. } 
\label{fig:OffsetVelApproach3}
\end{figure}

Finally, in the third approach, we compare the CO and \HI\ velocities point by point along the main axis of the Perseus cloud. In this approach, we account for velocity gradients~\citep{Ridgeetal2006, ImaraBlitz2011} along the cloud and find that the \HI\ emission velocities associated with this cloud tend to be smaller than the CO in absolute value, section by section and point by point. To ensure that our analysis is not impacted by velocity gradients or the uncertainties quoted in the first approach, we find \vcoHi\ for a number of points along the cloud (as illustrated in Figure~\ref{fig:OffsetVelApproach3}). In this final approach, we make sure that all the considered points have the exact same sky coordinates. The peak CO and HI LSR velocities for these points are shown with red and blue colors in Figure~\ref{fig:OffsetVelApproach3}, respectively, and yield an average CO velocity of  $1.4 \pm 0.6$\,km\,s$^{-1}$ in the co-moving \HI\ frame (where the 0.6 value is the standard deviation of the offsets in the peak velocities).  
Therefore, we conclude that for Perseus the average CO velocity in the co-moving HI frame (\vcoHi ) is positive (pointing away from us). This implies that, along the line of sight, the CO cloud is moving away from us faster than the \HI\ gas. We can also consider information about the Perseus cloud's surroundings and particularly presence of bubbles in this region. 

\subsection{Shock-cloud-interaction model}
\label{sec:SCIConsistency}

Knowing the \vcoHi\ and the GMF directions, we can make a prediction for the \blos\ reversal orientation across the cloud based on the SCI model, and compare it with our reconstructed 3D magnetic field and the \blos\ observations. For example, if the orientation of the GMF (as the initial magnetic fields in the region) and the filament main axis is similar to that of Figure~\ref{fig:GMFPers} (where the GMF unit vectors have a perpendicular component to the cloud axis), \vcoHi\ toward  us would indicate a convex arc-shaped magnetic field morphology, and \vcoHi\ away from us would result in a concave arc-shaped magnetic field morphology from our vantage point. Since the GMF vectors are pointing toward $-\hat{\ell}$ direction, we would expect a convex field to result in \blos\ toward us on the east side, and away from us on the west side. We would expect the opposite from a concave field: toward us on the west side, and away from us on the east. 

Therefore, the positive CO velocity in the co-moving \HI\ frame indicates that according to the SCI model, the GMF field lines ($-\hat{\ell}$ direction) should be bent and concave from our point of view.  
This means that the \blos\ observations should result in the fields pointing away from us on the south-east side of the cloud and toward us on the north-west side, which is consistent with the \blos\ observations of \citet{Tahanietal2018}, as shown in Figure~\ref{fig:PerseusBlosBpos}, and our reconstructed 3D field morphology, as depicted in Figure~\ref{fig:3DBPerseus}. This consistency indicates an even stronger likelihood for the arc-shaped magnetic field morphology.  

While the arc-shaped morphology illustrated in Figure~\ref{fig:3DBPerseus} is consistent with the predictions of the SCI model, we note that the morphology might also initially seem consistent with a scenario where the recurrent expansions of the Per-Tau shell have bent and ordered the field lines onto the shell. Examples of this, where the field lines take the shape of the shell and become tangential to the bubble (as the result of being pushed away) have been seen observationally~\citep[e.g.,][Tahani et al. in prep]{KothesBrown2009, Arzoumanianetal2021} and theoretically~\citep{KimOstriker2015, Krumholz2017}. 
If the Perseus cloud's arc-shaped magnetic field morphology is only as the result of tangential field lines to the bubble, then 1) the \blos\ reversal should be associated with the entire bubble and not specifically associated with the Perseus cloud, and 2) the \blos\ reversal should be relatively weak and not observable. However, the \blos\ observations of \citet{Tahanietal2018} suggest that the bending is directly associated with the Perseus cloud and does not continue beyond the Perseus cloud.  

Moreover, as discussed earlier, CO velocities have a higher magnitude in the direction away from us compared to the \HI\ velocities. If the cloud has formed by multi-compressions exerted only in the direction of the Per-Tau expansion, the CO velocities should naturally be slower than \HI , since in each expansion the bubble can push the \HI\ gas further.  
Therefore, the direct association of the observed \blos\ reversal with the Perseus cloud, as well as the positive \vcoHi , suggest that in addition to the Per-Tau bubble, the cloud has likely been influenced by another structure (pushing the cloud in the opposite direction to the Per-Tau bubble). This interaction has resulted in sharper bending of the magnetic field lines associated with the Perseus cloud and has pushed the \HI\ gas toward us, causing the faster CO velocities in the direction away from us.

To explore this, we investigate the available Planck thermal dust emission\footnote{Planck data release 2:  \url{https://wiki.cosmos.esa.int/planck-legacy-archive/index.php/Main_Page}}, \HI~\citep{HI4PICollaboration2016}, H$\alpha$~\citep{Finkbeiner2003}, 3D dust~\citep{Greenetal2019}, and CO composite maps~\citep{Dameetal2001}, which hint at the presence of a shell-like structure  at the far (toward the north) side of the Perseus cloud. This structure  has likely interacted with the Perseus cloud and is located on the far edge of the Per-Tau bubble.   
At the location of this structure, the \HI\ velocities are mostly negative ($-5$ to $1.2$\,km\,s$^{-1}$)  and point toward us, particularly in the co-moving frame of the Perseus cloud. 
This structure has likely pushed the surrounding HI region of the Perseus cloud toward us,  
resulting in a sharper bending of the field lines and the observed \vcoHi .

We refer to this shell-like structure as the Per2 shell or structure, which is consistent with the available 3D dust observations~\citep[][see their Figure 1]{Greenetal2019, Zuckeretal2021}.
The Per2 structure can be approximated as a circle in the plane of the sky centered at ($\ell = 157^{\circ}$, $ b = - 18^{\circ}$) with a radius of  roughly $5^{\circ}$, as shown in Figure \ref{fig:MultiWaveObsPerseus}.   
Figure~\ref{fig:MultiWaveObsPerseus} illustrates the Per-Tau and the Per2 shells in white dash-dotted and green dashed circles, respectively. The blue contours show the Perseus, California, and Taurus molecular clouds in the region observed by Herschel~\citep{Pilbrattetal2010}. As shown in the top-left panel, the Per2 shell is not identifiable in H$\alpha$ emission \citep{Finkbeiner2003}, since it is located behind the Per-Tau bubble. The Per2 shell-like structure is visible in Planck thermal dust observations, at $353$\,GHz and $857$\,GHz (as illustrated in the top right panel of Figure~\ref{fig:MultiWaveObsPerseus}), the CO composite survey of \citet{Dameetal2001}, and the HI4PI \HI\ velocity cube. 
Additional observations should be carried out in future studies to further confirm the presence of the Per2 bubble and its location. 

\begin{figure*}[htbp]
\centering
\includegraphics[scale=0.3, trim={2cm 0cm 1cm 1cm},clip]{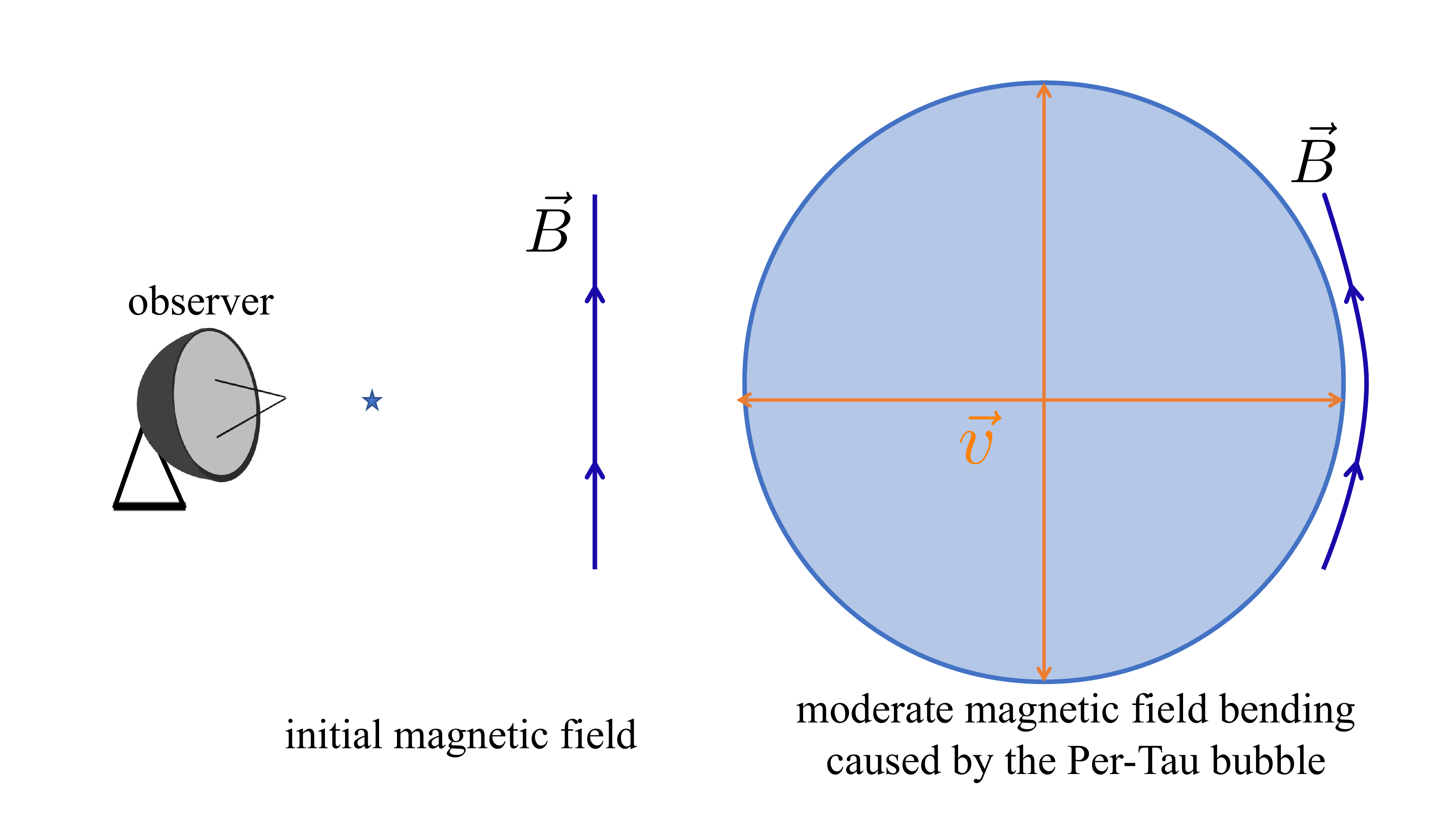}\\
\includegraphics[scale=0.3, trim={0cm 0cm 0cm 1cm},clip]{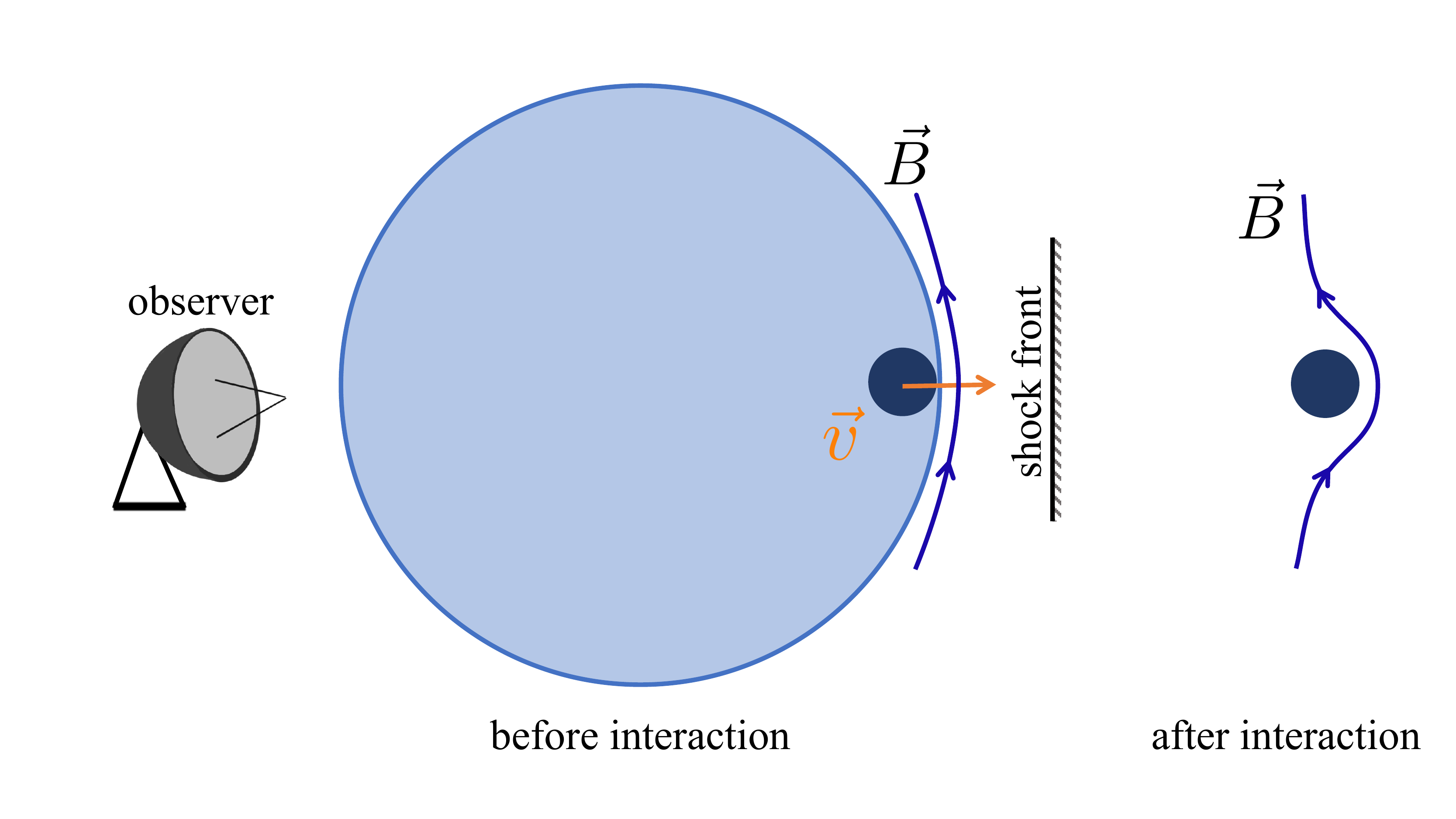}
\caption{Formation of the arc-shaped magnetic field morphology in the Perseus cloud. The large light blue circles depict the Per-Tau bubble (created by recurrent supernovae, represented by the star sign). 
The small dark blue circles depict an end-on view of the Perseus filamentary molecular cloud. \textbf{Top panel}: The straight \vec{B} on the left shows the direction of the initial magnetic field before the expansion of the Per-Tau bubble into the region. The curved \vec{B} line on the right illustrates the large-scale mild bending of the magnetic field morphology after the expansion. \textbf{Lower panel}: Formation of the arc-shaped magnetic field morphology around the Perseus molecular cloud due to interaction with the Per2 bubble, as described in the SCI model. The blue \vec{B} lines show the field lines before and after interaction with the Per2 structure on the left and right, respectively.}
\label{fig:SCIModifiedModel}
\end{figure*}

We suggest that the arc-shaped morphology associated with the Perseus cloud has been shaped by both the Per-Tau and Per2 bubbles: initially the multi-expansion (due to recurrent supernovae) of the Per-Tau bubble into the region can bend the component of the initial magnetic field that is perpendicular to the propagation direction, resulting in a tangential magnetic field morphology (a mild bending) as shown in the left panel of Figure~\ref{fig:SCIModifiedModel}~\citep[][see their Figure 16]{KimOstriker2015}. Subsequently, interaction with the Per2 structure pushes the surroundings of the cloud in the propagation direction of the Per2 shell at the Perseus cloud location (opposite to the Per-Tau bubble propagation direction), resulting in a sharper bending and the arc-shaped magnetic field morphology, as illustrated in the right panel of Figure~\ref{fig:SCIModifiedModel}.

\subsection{Other cloud-formation scenarios}

We note that simulations by \citet{LiKlein2019} find a bent (arc-shaped) magnetic field morphology in their filament formation model using a converging flow scenario (which is similar to the SCI model discussed in Section \ref{intro}). In their simulations, \citet{LiKlein2019} create a $\sim 4.4$\,pc-long filament, starting from a homogeneous box, with a sonic Mach number of $\mathcal{M}_s = 10$, a relatively strong initial magnetic field $\mathcal{M}_A = 1.62$, and a mean magnetic field strength of 20\,$\mu G$. \citet{LiKlein2019} find a long filament forming at the location of converging flows, with bent magnetic field lines. If initial magnetic field strengths are small, these field lines will be completely perturbed not following any large scale morphology. However, with strong magnetic fields, they find that the cloud retains a memory of the large-scale initial field morphology, even though smaller scale perturbations in the field lines are detected. 

Moreover, a bent magnetic field morphology might potentially be seen in simulations of~\citet[][see their Figure 10]{BastienBanerjee2015}, where they consider a converging flow scenario with 60$^{\circ}$ angle with respect to the filament (making it a similar case as the SCI model). More cloud formation models (other than the SCI model) could potentially predict arc-shaped magnetic field morphologies.  
We note that the Perseus cloud with $\ell \sim -20^{\circ}$ and a distance of $\sim 300$\,pc is off the disk and might have a different formation mechanism compared to the clouds located within the disk.  
Observations of many more molecular clouds at different locations within the Galaxy will allow us to more comprehensively examine the cloud-formation model predictions. 
This study calls for more predictions to be made using various cloud-formation models for a more comprehensive comparison with observations. 

\section{Discussion: reconstructed 3D magnetic morphology and its consistency with quantitative analyses}
\label{discussion}

In this section, we further discuss the reconstructed 3D magnetic field morphology, which incorporated 3D Galactic magnetic field models, as well as available \blos\ and \bperp\ observations. We investigate the influence of variations in the Galactic magnetic field models on the reconstructed field morphology and the consistency of this 3D magnetic field morphology with quantitative data.

\subsection{The 3D magnetic field morphology}

The coherent \blos\ and \bperp\ observations of the Perseus cloud, as well as the consistency of the \bperp\ with the GMF vectors projected on the plane of the sky, provide a simple case for reconstructing the 3D magnetic fields in this region. We note that the cloud is located toward the anti-Galactic-center direction and at high latitudes ($\ell \simeq 160^{\circ}, b \simeq -20^{\circ}$), resulting in less line-of-sight confusion compared to other clouds in the Galaxy. 

The consistency of our 3D magnetic field morphology with the cloud-formation model predictions further supports our reconstructed 3D magnetic morphology. This 3D morphology is also consistent with the effects of bubbles on the Perseus cloud: a) the Per-Tau bubble has caused a mild and large-scale bending in the field lines, and b) further interaction with the Perseus cloud's surroundings has pushed the \HI\ gas toward us and caused further field bending, resulting in the \blos\ reversal that is directly associated with the Perseus cloud. The location of the \blos\ observations and their error-weighted strengths indicate that the \blos\ reversal is directly associated with the cloud, caused by the interaction of the field lines with their surroundings. 

While we have reconstructed the most probable and natural 3D magnetic field morphology in the Perseus cloud and presented strong evidence for this 3D field, this study does not rule out other geometries completely. 
Future high-resolution multi-wavelength observations will allow us to more precisely and accurately determine the clouds' 3D magnetic field morphologies.

\subsection{Galactic magnetic field}
In Section~\ref{GMF}, we used the JF12 study to estimate the coherent component of the Galactic magnetic field, serving as an approximation for the initial magnetic field of the region in large-scales (before interactions with the surrounding bubble).  We found that our modeled GMF is perpendicular to the Perseus cloud and points from the southeast of the cloud to its northwest as shown in Figure~\ref{fig:GMFPers}. We also found that our modeled GMF vectors follow the same large-scale orientation that the Planck \bperp\ observations obtained in this region, indicating that the cloud retains a memory of the GMF. Since this GMF is approximately perpendicular to the Perseus cloud's axis, even large variations in the coherent GMF models (introduced by different parameters), would not change the overall 3D magnetic field morphology reconstructed in this study or the consistency that we see with the SCI model predictions.

\subsection{Age comparisons}
\label{sec:quantitatives}
Prior to identifying the Per-Tau bubble by~\citet{Bialyetal2021}, some studies identified a large sphere (\HI\  super-shell) associated with the Perseus cloud known as the Per OB2 association encompassing a longitude range from $\ell = 150^{\circ}$ to 180$^{\circ}$ and a latitude range of $b = -30^{\circ}$ to 0$^{\circ}$\citep{Ballyetal2008}. The age of the Per OB2 sphere was estimated to be from $\sim 6$ to 15\,Myr \citep{Blaauw1952, BorgmanBlaauw1964, Blaauw1964, Zeeuwetal1999, MeynetMaeder2000, Ballyetal2008}. The Per OB2 association encompasses a number of clusters and young stars. The Perseus cloud, itself, contains two clusters of IC 348 and NGC 1333, with approximate ages of 2\,Myr~\citep{Luhmanetal2003} and less than 1\,Myr~\citep{Ladaetal1996,
Wilkingetal2004}, respectively. 

Due to a younger age estimated for the Perseus cloud (1-5\,Myr), \citet{Ballyetal2008} suggest that the Per OB2 association has likely triggered the star formation in the Perseus molecular cloud. The Per-Tau shell identified in \citet{Bialyetal2021} is slightly larger than the Per OB2 sphere, and has a centroid location of ($\ell = 161.1^{\circ}$, $b = -22.7^{\circ}$) and a distance of 218\,pc, encompassing a longitude range of $\ell = 140^{\circ}$ to 185$^{\circ}$ and latitude range of $b = -40^{\circ}$ to 0$^{\circ}$. The Per OB2 and Per-Tau bubbles likely represent one bubble. \citet{Bialyetal2021} estimate the age of the Per-Tau shell to be around 6 - 22\,Myr, and suggest that multiple supernovae compressions from this shell have formed the Perseus molecular cloud.  

IC 348, which is located on the eastern side of the cloud and closer to the HI shell identified in previous studies~\citep{Shimajirietal2019, Doietal2021}, has greater stellar ages compared to the western side of the cloud, and was likely the first region in the Perseus cloud to undergo star formation~\citep{Bally2008}. We note that this eastern side of the cloud has larger distances from us compared to its western side~\citep{Zuckeretal2021}, which is due to its orientation on the Per-Tau shell identified by~\citet[][see their Figure 2]{Bialyetal2021}. In addition to different stellar ages across the Perseus cloud, studies suggest that there is a spread in the stellar ages found in IC 348 itself~\citep[e.g.,][]{Luhmanetal1998, Muenchetal2003}, which indicates that the region has likely undergone two incidents of star formation~\citep{Ballyetal2008}. The scenario illustrated in Figure~\ref{fig:SCIModifiedModel} can explain the presence of two separate populations of young stars, with different ages, through two epochs of star formation due to interaction with different bubbles in this region; one caused by multiple compressions by the Per-Tau bubble, and the other by interaction with the Per2 structure. 

\subsection{Pressure balance and multi-stage formation of arc-shaped magnetic fields}

\citet{Bialyetal2021} suggest that multiple supernovae have resulted in the Per-Tau shell and the formation of the Perseus molecular cloud. Using estimates of temperature and magnetic field strengths, we can find the gas and magnetic field pressures associated with the Perseus cloud at different evolutionary stages.  For this purpose, we use the following equations:
\begin{equation}
\begin{aligned}
&P_B = \frac{B^2}{8 \pi} [cgs],\\
&P_{\text{gas}} = nk_bT,
\end{aligned}
\end{equation}
where $P_B$, $B$, $P_{\text{gas}}$, $n$, $k_B$, and $T$ are the magnetic pressure, total strength of magnetic field, gas pressure, particle volume density, Boltzmann constant, and temperature, respectively. 

Before the formation of the Perseus cloud, we use an estimate of 5\,$\mu$G for the GMF strength, particle density of 100\,cm$^{-3}$, and an approximate temperature value of a few hundred kelvin ($\sim 500$\,K). We find a magnetic field and gas pressures of $1\times 10^{-12}$\,bar and $7\times 10^{-12}$\,bar, respectively. Therefore, magnetic field lines are initially not strong enough to resist the gas pressure and are easily bent and ordered or re-shaped by the bubble expansion~\citep{KimOstriker2015}. We note that the mass accumulation along with the bubble expansion result in higher magnetic field strengths on the shell. 
However, the external gas pressure (during the initial expansions) would still be higher than the magnetic pressure and therefore can further bend the magnetic field lines resulting in more ordered tangential field lines to the bubble in subsequent expansions. 

Simulations by~\citet{LiKlein2019} suggest that in the presence of strong fields $\mathcal{M}_A \simeq 1 $, magnetic field lines can coherently bend around a formed filamentary molecular cloud as the result of converging flows, while for $\mathcal{M}_A \simeq 10$, a complete perturbation of magnetic fields should be expected (and no coherent magnetic field morphology). We find the Alfv\'en mach number using the following equation:
\begin{equation}
\mathcal{M}_A = \frac{\sigma_{\varv}}{\varv_A},
\end{equation}
where $\sigma_{\varv}$ is the non-thermal velocity dispersion and $\varv_A$ is the Alfv\'en wave group velocity. The Alfv\'en velocity can be obtained using
\begin{equation}
\varv_A = \frac{B}{\sqrt{4 \pi \rho}} [cgs],
\end{equation}
where B is the strength of magnetic field and $\rho$ is the volume density. Using a velocity dispersion of 0.6\,km\,s$^{-1}$ as obtained for the Perseus cloud~\citep{Ponetal2014}, GMF strength of 5\,$\mu$G, and particle density of 100\,cm$^{-3}$, we estimate the initial $\mathcal{M}_A$ of around $0.5$, which is 2 orders of magnitude lower than the $\mathcal{M}_A$ values that could result in perturbed field morphologies (and ensures that the field lines should stay coherent and retain a memory of the initial ones). Higher $\sigma_{\varv}$ values (up to 6\,km\,s$^{-1}$) would still result in $\mathcal{M}_A$ values smaller than what could indicate a completely perturbed field morphology. Moreover, we note that recurrent supernovae  at the location of the Perseus cloud have likely enhanced the field strengths associated with this cloud, resulting in lower $\mathcal{M}_A$ numbers. 

To calculate the magnetic pressure of the Perseus cloud at its present state, we use the \blos\ and \bperp\ strengths that were previously obtained using Faraday rotation~\citep{Tahanietal2018}, Zeeman measurements~\citep{Crutcher1998, Crutcheretal1994, CrutcherTroland2000, Trolandetal2008, Thompsonetal2019}, and Davis Chandrasekhar Fermi method~\citep[DCF;][]{MatthewsWilson2002, Coudeetal2019}. The Zeeman and DCF observations mostly fall on the B1 region of the Perseus cloud. The widely-cited Zeeman magnetic field strength for B1 is $27 \pm 4\,\mu$G \citep{Goodmanetal1989}. Using a combination of available Zeeman observations for the \blos\ strength and the Davis-Chandrasekhar-Fermi (DCF) analysis for the \bperp\ strength, \citet{MatthewsWilson2002} and \citet{Coudeetal2019} estimate the total strength of the magnetic field in B1 ($\sim100\,\mu$G).  
Moreover, we find the error-weighted average of the \blos\ strengths obtained by \citet{Tahanietal2018} to be $136\,\mu$G and $-90\,\mu$G for the magnetic field data that point toward us and  away from us, respectively.

Considering these magnetic field strengths, we use an estimate of 100\,$\mu$G (order of magnitude) for the present-state magnetic field of the Perseus cloud. We find that the magnetic field pressure has a value an order of magnitude higher than the gas pressure, using particle densities of $10^3$ to $10^4$\,cm$^{-3}$ and temperature of 100\,K~\citep{Ponetal2014} to 12\,K, in regions from the cloud surroundings to the dense parts of the filament.  We also estimate the $\mathcal{M}_A$ of ~0.1 using 0.6\,km\,s$^{-1}$ for the velocity dispersion~\citep{Ponetal2014}. Therefore, the magnetic fields are strong enough to stay coherent and conserve their overall morphology. 

The presence of the Per2 shell-like structure can explain these strong magnetic fields (particularly along the line of sight) and their arc-shaped morphology. The enhanced pressure exerted by the Per2 bubble can bend the field lines further resulting in an arc-shaped field morphology and strong field strengths ($\sim 100\,\mu$G) directly associated with the Perseus filamentary structure. 

In summary, the a) lower magnetic field pressure compared to the gas pressure at initial stages (Per-Tau initial expansions), b) presence of coherent \bperp\ Planck observations c) presence of coherent \blos\ morphology, d) similar orientation of \bperp\ and GMF vectors, and e) the relatively-low initial $\mathcal{M}_A$ value strongly suggest that the initial bubble expansions have likely ordered and bent the field lines (instead of completely distorting them). This initial field bending, as shown in the left panel of Figure~\ref{fig:SCIModifiedModel}, is consistent with other observational~\citep[e.g.,][]{KothesBrown2009} and theoretical~\citep[e.g.,][]{KimOstriker2015} evidence and is similar to what is seen for \HII\ regions~\citep[e.g.,][Tahani et al. sub.]{Arzoumanianetal2021}, where the field lines are tangential to bubble edges. 
This is the first stage of the formation of the arc-shaped magnetic field morphology associated with the Perseus molecular cloud. 

Subsequently, the Per2 structure can result in a sharp bending of the field lines directly associated with the Perseus molecular cloud, as illustrated in the right panel of Figure~\ref{fig:SCIModifiedModel}.  Compression caused by the interaction with the Per2 structure, and subsequent matter flowing along the field lines and accumulating  on the filament, may result in self-gravity dominating the cloud's evolution and triggering a new sequence of star formation. It can also result in further enhancement of \blos\ strengths associated with the surroundings of the Perseus molecular cloud due to the sharper and more ordered field line bending. The a) spread in the stellar age population, b) higher CO velocities (compared to \HI\ with their directions pointing away from us), c) \blos\ reversal directly and sharply associated with the Perseus filamentary structure, d) presence of strong magnetic fields ($\sim 100\,\mu$G), and e) evidence of a shell-like structure present in the \HI , CO, and dust maps suggest that another structure/shell in the region (Per2) has likely interacted with the cloud and bent the field lines around it. 

\section{Summary and conclusions}
\label{sec:summary}
We reconstructed the 3D magnetic field morphology associated with the Perseus molecular cloud, using estimates for coherent Galactic magnetic fields~\citep{JanssonFarrar2012} and recent \blos\ observations~\citep{Tahanietal2018}. This field morphology is a concave arc shape from our point of view and points from southeast to the northwest of the cloud, in the $-\hat{l}$ direction (toward decreasing Galactic longitude).  While different studies~\citep[e.g.,][]{Clarketal2014, GonzalezLazarian2017, PattleFissel2019} provide powerful techniques to probe the plane-of-sky magnetic fields, to our knowledge this is the first time that the direction of plane-of-sky magnetic fields (of molecular clouds, scales of $\sim 10$-100\,pc) has been mapped. 

This field morphology retains a memory of Galactic magnetic fields and is an approximation, which neglects the small-scale magnetic field variations and fluctuations. Therefore, if there are fainter clouds in the foreground of the Perseus MC~\citep[e.g.,][]{UngerechtsThaddeus1987}, the magnetic fields of these fainter clouds do not interfere with this approximate large-scale arc-shaped magnetic field morphology. 

This arc-shaped morphology is consistent with the cloud-formation predictions of the shock-cloud-interaction model~\citep{Inutsukaetal2015, Inoueetal2018}. 
We explored the consistency of the shock-cloud-interaction model with
observational data, adding velocity observations and assuming that these velocity observations retain a memory of the most recent interaction. We studied the GMF direction, as well as the observed \HI\ and CO velocities and made predictions for the \blos\ reversal orientation associated with the Perseus molecular cloud based on the shock-cloud-interaction model. These predictions were consistent with the previous \blos\ observations~\citep{Tahanietal2018} and our reconstructed 3D magnetic field. 

In addition to our finding, some studies~\citep{Ballyetal2008, Doietal2020, Bialyetal2021} suggest that the Perseus molecular cloud has formed as the result of bubbles (shock-cloud-interaction) in this region. We suggest that the Per-Tau (Per OB2) bubble has first created a large-scale mild bending in the field lines. Subsequently, further interaction with the environment has bent the field lines, resulting in an arc-shaped magnetic field morphology directly associated with the Perseus cloud

While our observations are consistent with the shock-cloud-interaction model, this could be by chance and therefore exploring a large sample of clouds would be necessary. Upcoming rotation catalogs from surveys such as POSSUM~\citep{Gaensleretal2010} will allow for observing  \blos\ in many more clouds~\citep{Healdetal2020}.  These \blos\ observations can be used to reconstruct the 3D magnetic field morphology of more molecular clouds and to test the observations against predictions from different cloud formation models.

\begin{acknowledgements}
We thank the anonymous referee for their insightful comments that helped improve the results and the paper. MT is grateful for the helpful discussion with Pak-Shing Li. We have used \LaTeX, Python and its associated libraries including astropy~\citep{astropy} and plotly~\citep{plotly}, PyCharm, Jupyter notebook, SAO Image DS9, the Starlink~\citep{Starlink} software, the Hammurabi code, and Adobe Draw for this work. For our line integration convolution plot, we used a Python function originally written by Susan Clark. 
The Dunlap Institute is funded through an endowment established by the David Dunlap family and the University of Toronto. J.L.W. acknowledges the support of the Natural Sciences and Engineering Research Council of Canada (NSERC) through grant RGPIN-2015-05948, and of the Canada Research Chairs program. M.H. acknowledges funding from the European Research Council (ERC) under the European Union's Horizon 2020 research and innovation program (grant agreement No 772663).

\end{acknowledgements}

\bibliographystyle{aa} 
\bibliography{biblio}

\end{document}